\documentclass{aa}  
\usepackage{txfonts}
\usepackage{graphicx}
\usepackage{epstopdf}
\usepackage{natbib} 
\bibpunct{(}{)}{;}{a}{}{,} 
\begin{document}
\title{On the continuum intensity distribution of the solar photosphere}
\titlerunning{Intensity distribution of the solar photosphere}

\author{Sven Wedemeyer-B\"ohm\inst{1,2}\thanks{Marie Curie Intra-European Fellow of the European Commission}
\and 
Luc Rouppe van der Voort\inst{1}}

\offprints{sven.wedemeyer-bohm@astro.uio.no}

\institute{Institute of Theoretical Astrophysics, University of Oslo,
  P.O. Box 1029 Blindern, N-0315 Oslo, Norway
  \and
  Center of Mathematics for Applications (CMA), University of Oslo,
  Box 1053 Blindern, NÐ0316 Oslo, Norway
}

\date{Received 4 March 2009; accepted 5 June 2009}

\abstract
{For many years, there seemed to be significant differences between the 
continuum intensity distributions derived from observations and simulations 
of the solar photosphere.}
{In order to settle the discussion on these apparent discrepancies, we present a 
detailed comparison between simulations and seeing-free observations  
that takes into account the crucial influence of instrumental image 
degradation.}
{We use a set of images of quiet Sun granulation taken in the blue, green and red 
continuum bands of the Broadband Filter Imager of the Solar Optical Telescope (SOT) 
onboard Hinode. 
The images are deconvolved with Point Spread Functions (PSF) that account for non-ideal
contributions due to instrumental stray-light and imperfections.  
In addition, synthetic intensity images are degraded with the corresponding PSFs. 
The results are compared with respect to spatial power spectra, intensity histograms, 
and the centre-to-limb variation of the intensity contrast.}
{The intensity distribution of SOT granulation images is broadest for the blue continuum at 
disc-centre and narrows towards the limb and for longer wavelengths. 
The distributions are relatively symmetric close to the limb but exhibit a growing 
asymmetry towards disc-centre. 
The intensity contrast, which is connected to the width of the distribution, is found 
to be $(12.8 \pm 0.5)$\,\%, $(8.3 \pm 0.4)$\,\%, and $(6.2 \pm 0.2)$\,\% 
at disc-centre for blue, green, and red continuum, respectively.
Removing the influence of the PSF unveils much broader intensity distributions 
with a secondary component that is otherwise only visible as 
an asymmetry between the darker and brighter than average part of the distribution. 
The contrast values increase to 
$(26.7 \pm 1.3)$\,\%, $(19.4 \pm 1.4)$\,\%, and $(16.6 \pm 0.7)$\,\%
for blue, green, and red continuum, respectively.
The power spectral density of the images exhibits a pronounced peak 
at spatial scales characteristic for the granulation pattern and a steep decrease 
towards smaller scales. 
The observational findings like the absolute values and centre-to-limb variation 
of the intensity contrast, intensity histograms, and power spectral density  
are well matched with corresponding synthetic observables from three-dimensional 
radiation (magneto-)hydrodynamic simulations. 
}
{We conclude that the intensity contrast of the solar continuum intensity is higher 
than usually derived from ground-based observations and is well reproduced by modern 
radiation (magneto-)hydrodynamic models. 
Properly accounting for image degradation effects is of crucial importance for 
comparisons between observations and numerical models. 
} 

\keywords{Sun: photosphere; Radiative transfer}

\maketitle
%
\section{Introduction}

\begin{figure*}
  \includegraphics{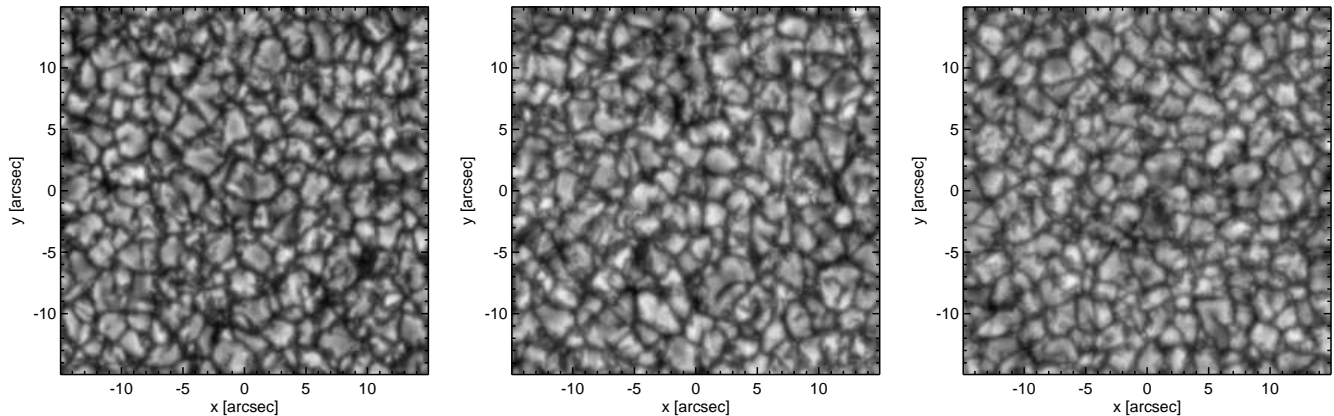}
  \caption{SOT/BFI observations at disc-centre: 
  Close-ups of intensity images in the blue (left), 
  green (middle), and red (right) continuum band. }
  \label{fig:sampleimages}
\end{figure*}

Modern numerical simulations of convection in the upper solar
atmosphere achieve a high degree of realism.
For example, in their landmark paper, \citet{2000SoPh..192...91S} 
show that their MHD simulations are very successful at reproducing a
number of observed properties of granulation.
Other remarkable examples are the excellent agreement between observed
and synthetic Fe line profiles \citep{2000A&A...359..729A} 
and between observed and simulated magnetic faculae 
\citep{2004ApJ...610L.137C, 2004ApJ...607L..59K, 2005A&A...430..691S}. 
As these diagnostic tests demonstrate the fidelity of the simulations, they
can be used to achieve a deeper understanding of the physical mechanisms
acting in the solar atmosphere and provide access to the un-observable
subsurface regions.
In recent years, extracting observables from simulations by synthesized 
spectra and intensity maps has grown into a powerful tool to understand and 
interpret observations \citep[see, e.g.,][and many more]{2005A&A...436L..27K, 
Langangen:2007lr}. 
Despite their success, there exist a number of diagnostic tests that
show considerable discrepancy.
One notorious example is the (rms) contrast of the continuum intensity
emerging from the low solar photosphere, also referred to as
granulation contrast or granular intensity fluctuation 
\citep[see, e.g.,][]{1975A&A....45..167D,2007MmSAI..78...71M}. 
Its value is straightforward to calculate and it is deeply connected
to the basic mechanisms of convection as it is sensitive to the
distribution of bright (hot) and dark (cool) surface structures.
It is therefore troubling that contrast values derived from
simulations do not agree with observed contrast values 
\citep[see, e.g.,][]{1984ssdp.conf..174N,2006ASPC..354....3K} 
with synthetic contrast values being usually significantly higher.
For example, \citet{2007ApJ...668..586U} compared G-band images of the quiet Sun.
They report contrasts of 14.1\,\% for the (reconstructed) observed image
and 21.5\,\% for the synthetic image from a model by
\citet{2006ApJ...642.1246S}.

Qualitatively, empirical studies show that the granulation contrast
decreases with increasing wavelength and decreases with increasing
distance from the disc-center.
\citet{1997A&A...325..819W} used the German Vacuum Tower Telescope
(VTT) to obtain granulation contrast at different heliocentric
positions at a wavelength of 550\,nm.
They derive $(13.5\,\pm\,1)$\,\% for $\mu = 1.0$ (disc-centre),
8\,-\,9\,\% at $\mu = 0.1$, and down to 3.8\,\% very close to the
solar limb at $\mu = 0.07$.
The wavelength dependence and center-to-limb variation of the contrast
provide a stringent diagnostic test for simulations as it connects to
the temperature stratification of the photosphere.

In observations many factors affect the contrast and most of them are
poorly understood.
Imperfections in the optics and straylight significantly reduce the
contrast but are very difficult to measure and correct for.
In addition, image degradation by the Earth atmosphere (seeing) has a
huge impact on the image quality and can at best only partly be
corrected for.
The determination of the centre-to-limb variation of the mean absolute 
continuum intensity already requires enormous efforts 
\citep[e.g.,][]{1984SoPh...90..205N}. 
Owing to these hardly controllable factors, a large range of contrast
values are reported in the literature. 
See  for example the review by \citet{1990ARA&A..28..263S} 
and more recently Table~2 by \citet{2000ApJ...538..940S}
and Fig.~1 by \citet{2008PhST..133a4016K}. 
Continuous advancement of observational methods, in particular the
development of adaptive optics and image reconstruction techniques,
lead to an increasing trend of the reported contrast values. 
These methods like for example speckle
\citep[e.g.,][]{1993A&A...268..374V, 1993PhDT.......155D, 
2008A&A...488..375W} 
and Phase Diversity/MOMFBD \citep{2005SoPh..228..191V} involve measurements of 
the effective point-spread-function (PSF).
Despite the high level of sophistication and impressive results, a
complete reconstruction (i.e., full removal of the combined effect of
seeing-induced degradation and instrumental aberrations) can never be
obtained.
Contributions to the PSF that are important for accurate photometry --
essential for the determination of the granulation contrast -- are
often hardly known.
This is in particular true for stray-light produced inside the optical
instrument \citep{1983SoPh...87..187M}.
Already a simple uniform intensity offset reduces the resulting
intensity contrast.
In most cases, however, the straylight contribution is likely
to be anisotropic and varying over the field of view (FOV).

With the advent of the Solar Optical Telescope
\citep[SOT,][]{2008SoPh..249..167T,
2008ASPC..397....5I,
2008SoPh..249..197S,
2008SoPh..249..221S}
onboard the Hinode spacecraft \citep{2007SoPh..243....3K}, the solar
community has now access to space observations that have sufficient
spatial resolution to resolve the scales that are important for an
accurate determination of the granulation contrast.
A space-born instrument has the obvious advantage of the absence of
seeing so that only the instrumental PSF or stray-light needs to be
considered.
\citet[][hereafter referred to as Paper~I]{2008A&A...487..399W}
derived a set of PSFs for the combined effect of the Broadband Filter
Imager (BFI) and the Optical Telescope Assembly (OTA) of SOT.
He concluded that there is a small but significant amount of
stray-light, although the instrument still operates close to the
diffraction limit.
The effect on the intensity distribution, including the granulation
contrast, could thus be significant.
This view is supported by \citet{2008A&A...484L..17D}, who find that
the difference in contrast between observations with the
Spectro-polarimeter (SP) of SOT and state-of-the-art numerical
simulations is strongly reduced, when taking into account instrumental
image degradation.

In this paper, we exploit the extensive Hinode data base that has been
built up through its successful operations.
Over time, the various observing programs have covered the quiet Sun
at a whole range of observing angles providing the data to facilitate
a robust statistical study of the wavelength dependence and
center-to-limb variation of granulation contrast.
Furthermore, we consider different independent numerical models. 
The paper is organised as follows: the observations are described in
Sect.~\ref{sec:obs}, followed by a description of the PSFs in Sect.~\ref{sec:psf} 
and the method of deconvolution in Sect.~\ref{sec:decon}.  
The numerical simulations and the intensity synthesis are introduced in 
Sect.~\ref{sec:synintens}. 
The observations and the simulations are compared in detail for disc-centre 
images in Sect.~\ref{sec:compwsim}, followed by an analysis of the 
centre-to-limb variation of the intensity distribution and the continuum 
intensity contrast in Sect.~\ref{sec:intdis}. 
Finally, the observations and simulations are compared in terms of 
spatial power spectral density in Sect.~\ref{sec:spatpower}.  
Discussion and conclusions can be found in Sects.~\ref{sec:discus} 
and~\ref{sec:conc}, respectively.

\section{Observations}

\subsection{SOT filtergrams}
\label{sec:obs}
The observations analysed in this study were obtained with the Broadband 
Filter Instrument (BFI) of the Solar Optical Telescope (SOT) onboard the Hinode 
satellite (see Sect.~1 for references) in the period from November 2006 until 
February 2008. 
Wide-band filtergrams in the blue, green and red continuum from various dates are 
considered (see Fig.~\ref{fig:sampleimages}). 
The nominal central wavelengths of these filters are 
450.45\,nm, 555.05\,nm, and 668.40\,nm, 
respectively. 
The transmission profiles of these filters, which are available in Solar Soft 
\citep{2000eaa..bookE3390F}, 
have FWHMs of 0.22\,nm, 0.27\,nm, and 0.31\,nm, respectively. 
As in Paper~I, the position of the FOV and individual subregions on the solar 
disc is here reduced to the heliocentric position \mbox{$\mu = \cos \theta$} with 
$\theta$ being the observing angle.
It is justified by the finding that the intensity profiles across the solar limb 
do not show noticeable differences between different positions along the limb (N/W/S/E)  
\citep[cf.][]{2002A&A...382..312L}. 
The FOV is 223\,\arcsec$\,\times\,$112\,\arcsec for images that use the whole detector and 
112\,\arcsec$\,\times\,$112\,\arcsec for images that use only half. 
The pixel scale of the detector of only 0\,\farcs054 is so much smaller than the 
spatial resolution of the optical system 
\citep[0\,\farcs2 - 0\,\farcs3, see, e.g.,][]{2008SoPh..249..221S} 
that images with $2 \times 2$ pixel-binning can be used next to unbinned images 
without causing significant differences in terms of contrast and intensity distribution. 
A total number of 584~images of quiet Sun granulation are analysed, covering 
heliocentric positions from disc-centre to the solar limb. 
The images were selected to contain quiet Sun only, active region
observations were excluded. 
Basic data processing (i.e., dark subtraction and flat-fielding) was
preformed using routines from the Hinode branch of Solar Soft.

\begin{figure} 
  \centering
  \includegraphics[width=7cm]{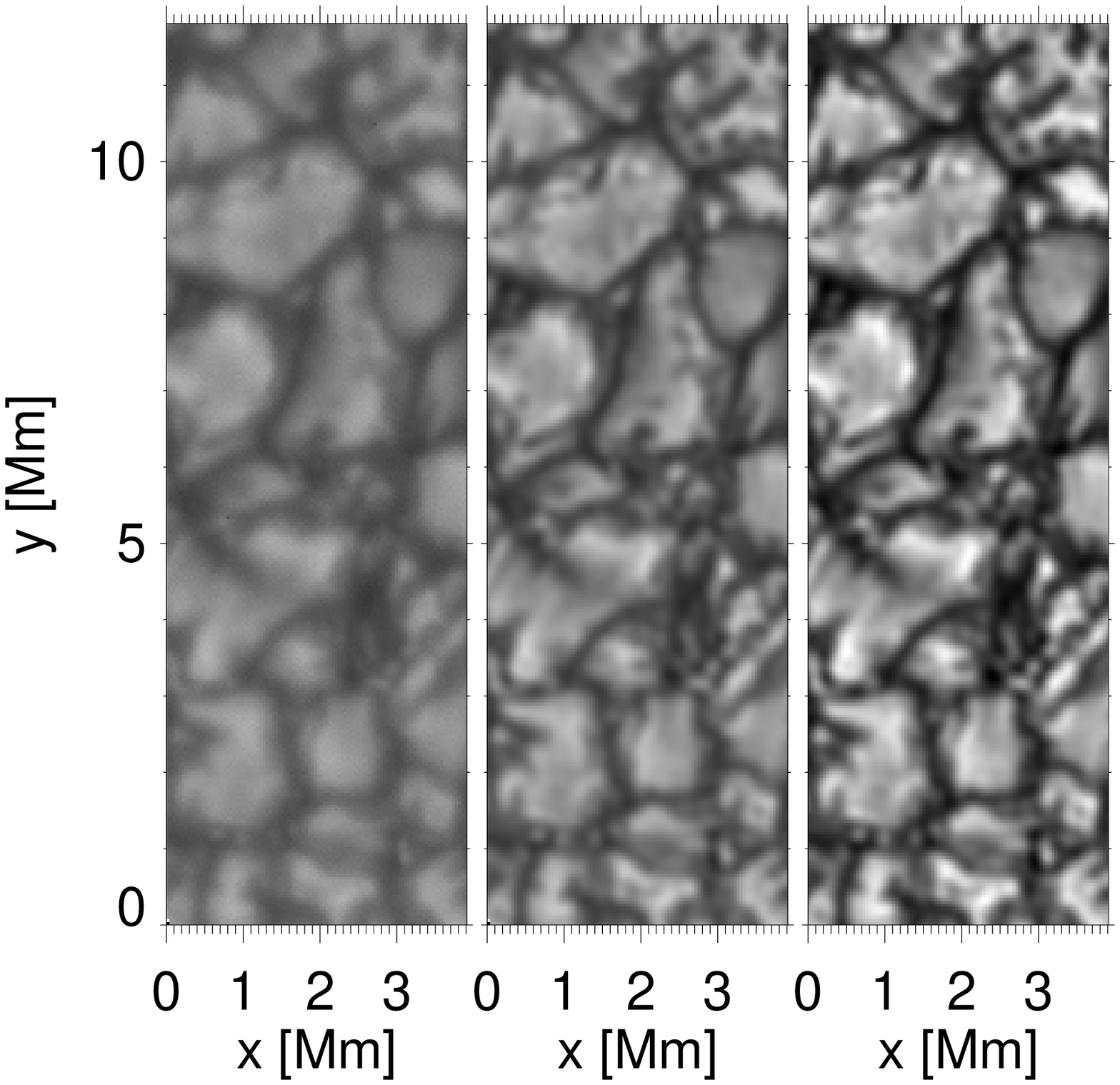}
  \includegraphics[width=7cm]{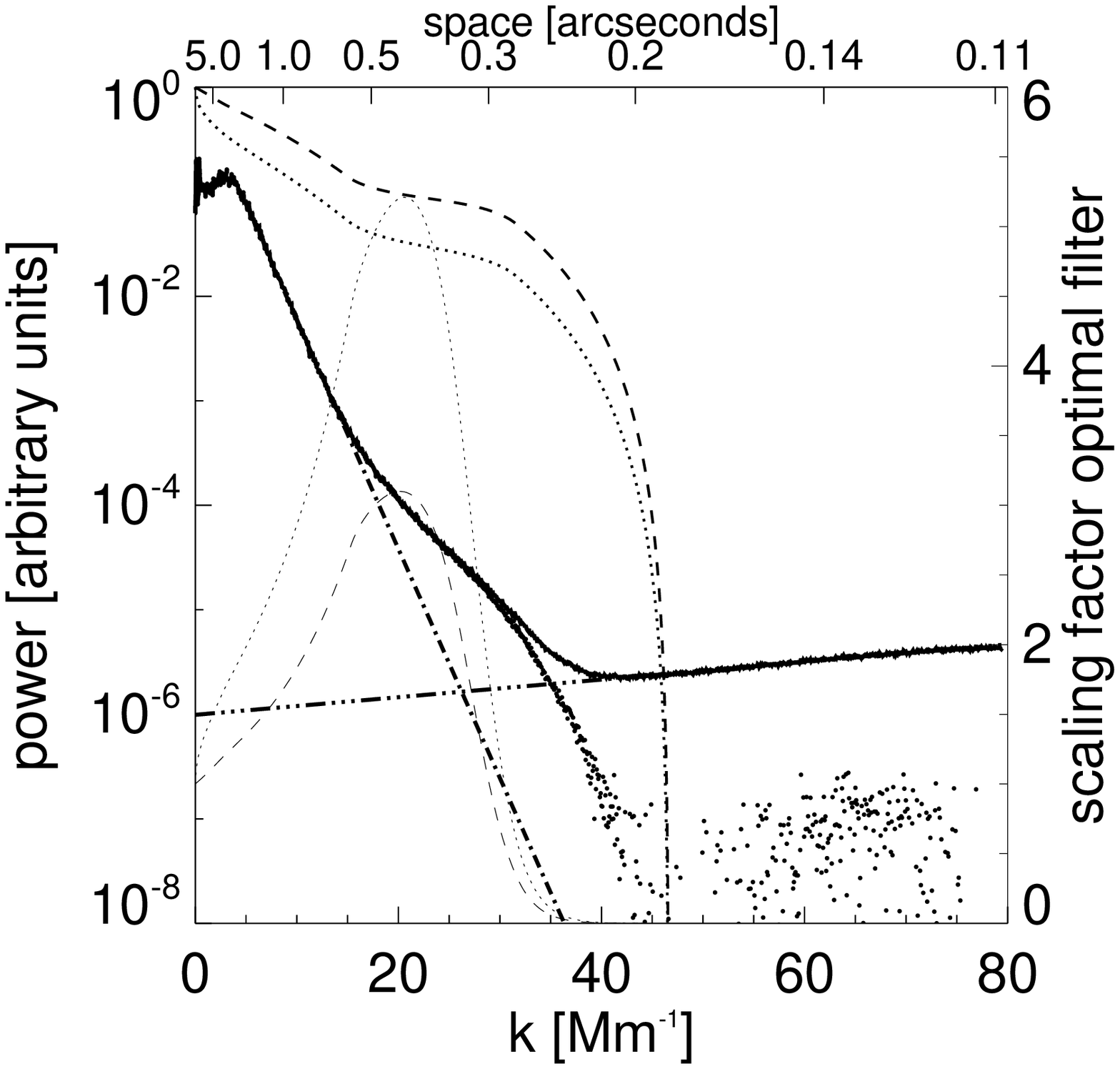}
  \caption{Illustration of deconvolution using optimal (Wiener)
  filtering. The top panel shows a detail of a blue continuum disc
  center image. The left image is the flatfielded image, the middle
  image after deconvolution using the ideal PSF, and the right image
  after deconvolution using the non-ideal PSF from Paper~I. All three
  images are scaled using the same intensity range: the minimum and
  maximum pixel values of the right panel. In the graph below, the thick
  solid line is the azimuthal average of the 2D power spectrum of a blue
  continuum disc center image. The triple-dot-dashed line is the noise
  model; the dots represent the power spectrum with the noise model 
  subtracted. 
  The dash-dot line is the model for the signal in the absence of
  noise. Lines in medium thickness: the dashed line is the ideal MTF, the
  dotted line the non-ideal MTF. Both have their cut-off at the
  diffraction limit $\lambda/D = 0\,\farcs19$. Thin lines: the dashed
  line is the optimal filter for the ideal PSF, the dotted line for the
  non-ideal PSF. Both lines relative to the y-axis scaling on the right
  side.} 
  \label{fig:wiener}
\end{figure}

\subsection{Point Spread Functions}
\label{sec:psf}

We use the point spread functions for the Solar Optical Telescope as they are 
calculated in Paper~I. 
There are individual ideal and non-ideal PSFs for the blue, green, and the red 
continuum band.  
The ideal PSFs represent the image degradation due to diffraction at the main 
aperture stop of the telescope, which has a diameter of 50\,cm.
The central obstruction and the spiders are taken into account. 
The non-ideal PSFs describe the combined effect of the Broadband Filter Imager and 
the Solar Optical Telescope.  
In contrast to the ideal PSFs, they account for contributions caused by 
imperfections and contamination of the optical components in the telescope
(e.g., impurity of lens material, dust on optical surfaces, microscopic scratches 
or micro-roughness) and stray light, which may originate, e.g., from reflection 
at baffles. 
The non-ideal PSFs are calculated by convolution of the ideal diffraction-limited 
PSFs and Voigt profiles, which are chosen as model for the non-ideal contribution. 
For each wavelength channel, such theoretical point spread functions were produced 
for a grid of the parameters $\sigma$ and $\gamma$, which control the shape of the 
non-ideal Voigt-like contribution. 
These PSFs were then applied to artificial images of an eclipse and a Mercury transit. 
In a next step, the grid of resulting artificial intensity profiles across the terminators 
were compared to SOT observations of the Mercury transit from November 2006 and the 
solar eclipses from 2007. 
For each observed image in the blue, green, and red continuum channels of the BFI, 
the optimum PSF is indicated by the best-fitting artificial profile, resulting 
in the parameters $[\sigma, \gamma]$ and a usually small residual intensity offset. 
As the non-ideal contributions depend on a number of factors, the individual images 
produce a range of possible parameter combinations rather than a single solution. 
Nevertheless, the stray-light contributions are found to be best matched with Voigt 
functions with the parameters 
$\sigma = 0\,\farcs008$ and $\gamma = 0\,\farcs004$, $0\,\farcs005$, and $0\,\farcs006$ 
for the blue, green, and red continuum channels, respectively.  

\subsection{Deconvolution}
\label{sec:decon}

The granulation images are deconvolved with (i)~the ideal (diffraction-limited) 
PSFs and (ii)~the non-ideal PSFs that are described in Sect.~\ref{sec:psf}.
We use all images at disc-centre for each wavelength channel (15~images 
each).  
The rectangular Hinode images were split into two square images for deconvolution in
order to facilitate a straightforward execution and interpretation
using standard FFT methods.  
After deconvolution, the average contrast
of both sides is used for the analysis.

The deconvolution was performed by means of an optimal (Wiener) filter
\citep[see, e.g., Chapter 5.8 of][]{gonzalez2008image_processing}.  
Figure~\ref{fig:wiener} illustrates various aspects of the construction of the 
optimal filter. 
The graph shows the azimuthal average of the 2D power spectal density 
(or ``power spectrum'', see Sect.~\ref{sec:spatpower}) of signal plus
noise as the solid black line. 
In the signal-dominated regime, the power spectrum has a peak followed by 
decreasing power for the intermediate range of increasing wavenumbers~$k$. 
In the noise-dominated regime, the power increases with increasing wavenumber 
for spatial scales smaller than the diffraction limit 
$\lambda/D < 0\,\farcs19$ (see the cut-off for the MTF). 

For the construction of the optimal filter, an assumption has to be
made for the power spectrum of the signal in the absence of noise 
\citep[see the discussion in Chapter 13.4 of][]{1992nrfa.book.....P}.  
To this end, a noise
model was constructed by extrapolating a fit to the noise tail into
the signal region (triple-dot-dashed line). 
The subtraction of the noise model from
the measured power spectrum is shown as black dots. The power spectrum
displays an excess in the form of a less steep power fall-off around
spatial scale 0\,\farcs3. A similar excess is seen in all power spectra
and becomes more pronounced for the green and red continuum
channels. Since we do not observe such excess in power spectra from
higher spatial resolution granulation images from the Swedish 1-m
Solar Telescope \citep[see, e.g.,][]{2004A&A...414..717R}
and the Dunn Solar Telescope \citep[see, e.g.,][]{2008A&A...488..375W}, 
we decided to
exclude this excess from the signal model. The origin of the excess is
unclear but we can exclude JPEG compression artifacts (see
Sect.~\ref{sec:jpeg}). 
The signal model is derived from an extrapolation of the
power spectrum to high wave numbers following the power law in the
[0\,\farcs6 -- 1\arcsec] range (dash-dotted line).

The azimuthal average of the optimal filter is also shown in the graph
of Fig.~\ref{fig:wiener} (thin lines, following the right y-axis scaling). 
The filter enhances ($>1$) for spatial scales $\geq 0\,\farcs27$ and suppresses noise for
smaller scales. Although the shape of the optimal filter is
significantly affected by in- or exclusion of the power excess in the
signal model (including the excess shifts enhancement of the filter
down to 0\,\farcs2), the contrast is affected only little: 0.1\% higher 
contrast\footnote{All contrast values and differences between them are 
given in percent points, notified by \% (see Sect.~\ref{sec:contrast}).}   
for the ideal PSF and 0.2\%  for the non-ideal PSF. 
This illustrates that the contribution of small spatial scales to the overall 
contrast is limited. 
That makes deconvolution by an optimal filter a robust procedure, where details of
the construction of the filter have only limited impact on the
contrast values.

While the deconvolution of disc-centre images produces rather robust results, 
the treatment of images at smaller $\mu$ is problematic. 
In particular, images close to the solar limb cover a significant range in $\mu$. 
Consequently, the properties of the stray light component vary significantly across
the FOV. 
The same applies to the noise in power spectra that are used for the construction for 
the Wiener filter. 
It is therefore not possible to construct a Wiener filter that is optimal for the whole 
FOV. 
Retreating to deconvolution of smaller image sections with small intrinsic variation in $\mu$
would cause systematic errors due to neglect of the far PSF wings. 
The contrast values for deconvolved images away from the disc-centre would suffer a 
pronounced spread and possibly little significance only. 
We therefore restrict the deconvolution to disc-centre images.  

The three images in Fig.~\ref{fig:wiener}  show the effect of the deconvolution. 
The left panel shows the flatfielded image, the middle panel after
deconvolution considering the ideal PSF, and the right panel after
considering the non-ideal PSF from paper I. All images are shown using
the same intensity scaling. The effect of noise suppression for the
two right panels is clear. 
The contrast is enhanced from 12.6\,\% for
the flatfielded image to 17.9\,\% for the image deconvolved with the
ideal PSF, and to 26.5\,\% after deconvolution with the non-ideal PSF.
See Table~\ref{tab:irmscomp} for statistics for all disc-centre images.

\begin{table}[b]
  \caption[h]{Numerical simulations used in this study.}
  \label{tab:sim}
  \centering
  \begin{tabular}[b]{l|llcc}
  \hline
  \hline
  &&\multicolumn{3}{c}{parameters}\\
  ID&reference&RT$^{\mathrm{a}}$&$\langle B_z \rangle$$^{\mathrm{b}}$&hor. ext.$^{\mathrm{c}}$\\
  &&&[G]&[Mm]\\
  \hline
  \multicolumn{4}{c}{CO$^5$BOLD simulations}\\
   \hline
   Ch0&Steffen (2007)              &non-gr.& 0&11.2\,$\times$\,11.2\\
   Ch1&Wedemeyer et al. (2004)     &grey   & 0& 5.6\,$\times$\,5.6\\
   Ch2&Wedemeyer et al. (2004)     &non-gr.& 0& 5.6\,$\times$\,5.6\\
   Cm &Schaffenberger et al. (2006)&grey   &10& 4.8\,$\times$\,4.8\\
   \hline
   \multicolumn{4}{c}{Stein~\& Nordlund simulations}\\
   \hline
   SNh&Stein \& Nordlund (2006)&non-gr.&  0&6.3\,$\times$\,6.3\\
   SNm&Stein \& Nordlund (2006)&non-gr.&250&6.3\,$\times$\,6.3\\
  \hline 
  \hline
  \end{tabular}
  \begin{list}{}{}
  \item[$^{\mathrm{a}}$] 
    For each model it is indicated if frequency-independent (``grey'') or 
    -dependent (``non-grey'') radiative transfer (RT) was considered. 
  \item[$^{\mathrm{b}}$] 
  magnetic field $\langle B_z \rangle$
  \item[$^{\mathrm{c}}$] 
  horizontal extent of the computational domain
  \end{list}
\end{table}

\section{Synthetic intensity}
\label{sec:synintens}

\subsection{Numerical simulations}
\label{sec:sim}

An overview over the models used here is given in Table~\ref{tab:sim}. 
Most of our analysis will be based on three snapshots 
of a recent 3D radiation hydrodynamic simulation by 
\citet{2007IAUS..239...36S}, which was computed with 
\mbox{CO$^5$BOLD}  
\citep{cobold, 2004A&A...414.1121W, 2008asd..soft...36F}. 
The model, which is identified as Ch0 hereafter, has a horizontal grid spacing 
of 28\,km and an extent of 11.2\,Mm\,$\times$\,11.2\,Mm. 
It reaches from $\sim -2400$\,km below the average height of optical depth 
$\tau_{500\,nm} = 1$ to $+750$\,km above. 
The simulation code \mbox{CO$^5$BOLD} employs a Riemann-type solver. 
Note that the effective spatial resolution of models produced with such a code 
is higher than for models calculated with a finite-differences solver at 
same grid spacing. 
The radiative transfer in the simulation uses a frequency-dependent 
(``non-grey'') long characteristics scheme with multi-group opacities with five bins. 
The latter are constructed from updated  MARCS  opacities 
\citep[see][and references therein]{2008A&A...486..951G}. 

For comparison, additional sets of synthetic images are calculated for 
\mbox{CO$^5$BOLD models} and simulations by \citet{1998ApJ...499..914S}. 
All sets consist of three snapshots each. 
The average contrast and standard variation for disc-centre images are given 
in Table~\ref{tab:irmscomp}.
The abbreviations Ch1 and Ch2 stand for the non-magnetic models by 
\citet{2004A&A...414.1121W} with frequency-independent (``grey'') and 
frequency-dependent (``non-grey'') radiative transfer, respectively.
Cm is also a \mbox{CO$^5$BOLD} model but with a weak magnetic field of 
$\langle B_z \rangle = 10$\,G \citep{2006ASPC..354..345S}. 
The models by Stein \& Nordlund only differ such that 
SNh has no magnetic field, whereas 
SNm has $\langle B_z \rangle = 250$\,G. 
A snapshot of the latter has already been used by, e.g., \citet{2004ApJ...610L.137C},  
\citet{2006A&A...449.1209L}, 
and by \citet{2007ApJ...668..586U}. 
See also \citet{2006ApJ...642.1246S}. 

\begin{figure*}[t]
  \sidecaption
  \includegraphics[width=12cm]{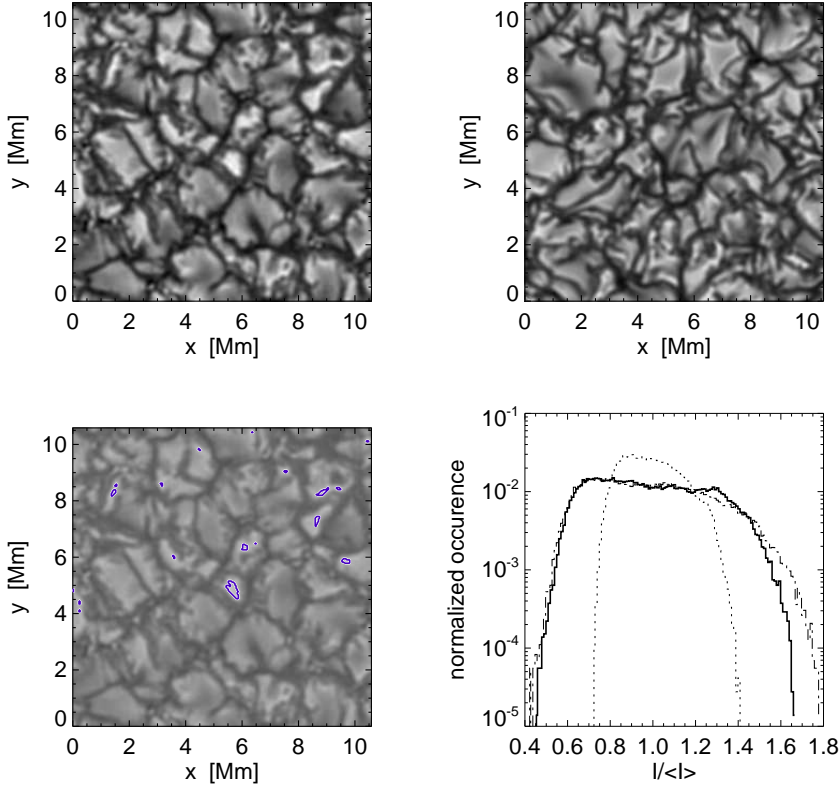}
  \caption{Comparison of synthetic and observed images for the blue 
  continuum channel. 
  \textit{Top left}: observed deconvolved image, 
  \textit{top right}: synthetic image rescaled to the same
   pixelsize as the observed image;     
  \textit{bottom left:} same as top left but with contours enclosing
  the pixels, which are brighter than in the synthetic image 
  ($I/\langle{}I\rangle > 1.66$). 
  The image is made fainter for better visibility of the contours. 
  \textit{Bottom right:} histograms for the intensity distributions:
  original observed (dotted), deconvolved (dot-dashed), and original
  synthetic intensity (thick solid). }
  \label{fig:comp_deconobs}
\end{figure*}

\begin{figure*}[t]
\sidecaption
\includegraphics[width=12cm]{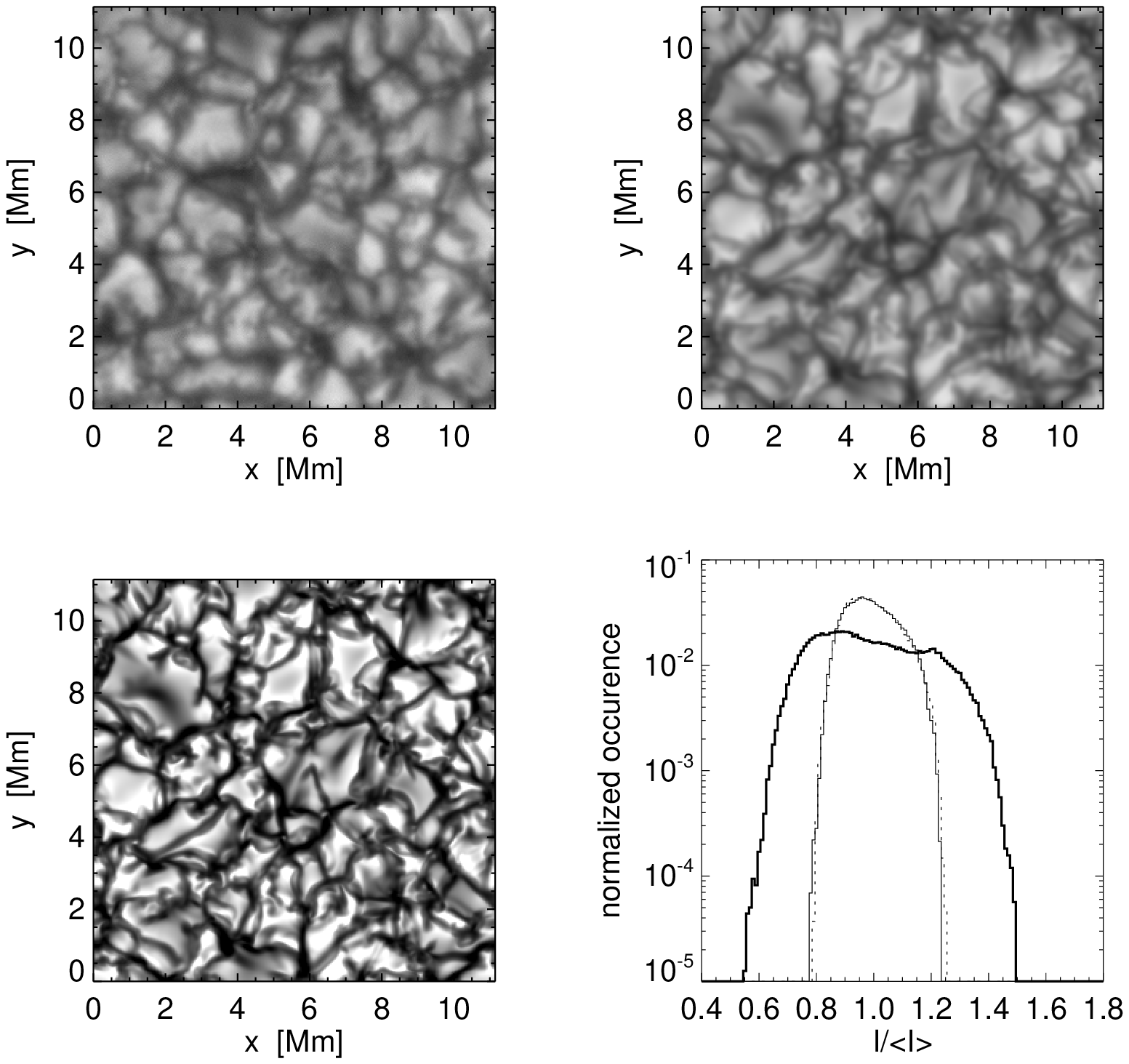}
\caption{Comparison of synthetic and observed images for the green continuum 
  channel. 
  \textit{Top left}:  original observed image, 
  \textit{top right}: degraded synthetic image rescaled to the same 
  pixelsize as the observed image;     
  \textit{bottom left:} original synthetic image; 
  \textit{bottom right:} histograms for the observed intensity
  (dotted), degraded synthetic intensity (thin solid), and original
  synthetic intensity (thick solid).} 
\label{fig:comp_degradsim}
\end{figure*}

\subsection{Intensity synthesis}

We use the spectrum synthesis code Linfor3D to produce synthetic intensity maps 
for the radiation \mbox{(magneto-)}hydrodynamic simulations listed in 
Table~\ref{tab:sim}. 
Linfor3D (see \mbox{http://www.aip.de/$\sim$mst/linfor3D$\_$main.html)}
is originally based on the Kiel code LINFOR/LINLTE.  
It solves the detailed radiative transfer in full 3D for an input atmosphere under 
the assumption of local thermodynamic equilibrium (LTE). 
The computations are performed on a numerical grid that has a higher 
resolution compared to the input model. 
For consistency reasons, the same opacities and element abundances are used 
in the numerical simulations and in Linfor3D (e.g., updated  MARCS  opacities 
for model Ch0, see Sect.~\ref{sec:sim}). 

The green and in particular the red continuum band contain only a few mostly weak 
spectral lines so that their relative influence on the radiation integrated over the
whole wavelength band is negligible. 
The blue continuum channel includes a spectral line of neutral iron (\ion{Fe}{I}) at 
450.483\,nm, which we explicitly take into account. 
The line parameters are taken from the Vienna Atomic Line Database \citep{1999A&AS..138..119K}.  
The influence of the line on the radiation integrated over the full blue continuum filter 
turns out to be negligible. 
The change in contrast would be well below 0.1 percent point with respect to pure continuum 
images and also the intensity histograms are barely affected. 

\subsection{Degradation of synthetic maps}

The PSFs introduced in Sect.~\ref{sec:psf} are used to simulate the image 
degradation caused by the SOT/BFI optics. 
Each synthetic intensity map is convolved with the non-ideal PSF of the 
corresponding wavelength. 
For the model Ch0, we also use the ideal PSFs and two different FOVs (see 
discussion in Sect.~\ref{sec:adduncert}).

\section{Comparison of observed and synthetic images}
\label{sec:compwsim}

We begin our comparison of the Hinode filtergrams and the numerical 
simulations on the basis of disc-centre images. 
First, the de-convolved observations are opposed to the original synthetic 
intensity maps from model Ch0 in Sect.~\ref{sec:comp_deconobs}. 
The original observations are then checked against the corresponding degraded 
synthetic maps in Sect.~\ref{sec:comp_degsyn}. 

\subsection{Synthetic images compared to deconvolved observations}
\label{sec:comp_deconobs}

The synthetic images have a lower contrast than the PSF de-convoluted SOT/BFI images  
but are still within the 1-$\sigma$ variation of the observed values in the blue 
and green continuum.
The differences are 1.7\,\%, 1.3\,\%, and 1.9\,\% for the three channels, 
respectively. 
The comparison is illustrated in Fig.~\ref{fig:comp_deconobs} for blue continuum 
images at disc-centre. 

In order to make a fair comparison, the synthetic image is filtered
with a low-pass filter. The filtering accounts for the fact that the
observations are taken with a telescope with finite
aperture. Consequently, there can be no higher spatial frequencies
present than given by the size of the aperture. In addition, the
observations contain noise at a level that affects the highest spatial
frequencies (the power spectrum of the image is dominated by noise at
frequencies higher than the diffraction limit). The low-pass filter was
constructed in a way such that the power spectrum of the filtered
synthetic image matches the power spectrum of the deconvolved
image. After convolution, the synthetic image was re-sampled to the
Hinode pixel scale. The observed image was selected to be as quiet as
possible, selecting a region with as few magnetic bright points as
possible. 
The top left image in Fig.~\ref{fig:comp_deconobs} shows the deconvolved observed
image, the top right image the filtered synthetic image. Both are
shown using the same intensity scaling. In a visual comparison of the
two images, one can distinguish the observed from the synthetic image by the presence of
a noise pattern that remains after application of the optimal filter. 
Other than this noise pattern, the two images are qualitatively
remarkably similar.

The lower right panel of Fig.~\ref{fig:comp_deconobs} shows the histograms of the intensity
values of the two images. In addition, the histogram of the
flatfielded observed image is shown (dashed line). Deconvolution by
the non-ideal PSF has a pronounced effect on the intensity
histogram. The histograms of the deconvolved observed and synthetic
images are very similar. It should be noted that the low-pass
filtering of the synthetic image has only limited effect on the
histogram. 
Both the synthetic images and the deconvolved observations exhibit an 
essentially ``double-peaked'' distribution,  
although the peak at $I/\langle{}I\rangle < 1$ is much more pronounced than 
the secondary peak at  $I/\langle{}I\rangle > 1$.
The latter can easily be obscured and only be noticable as asymmetry in the 
distribution (see Fig.~\ref{fig:sothist}). 
The distribution can actually be decomposed in a bright granular and a 
dark intergranular component for the simulations, when one considers 
the sign of the flow velocity (downflow/upflow). 
This property has been known for some time 
\citep[see, e.g., Fig.~3 in][]{2000SoPh..192...91S}. 
A remarkable difference between the two histograms, however, can be
seen at high intensity values. These high pixel values (160 in total,
or 0.2\%) can be found as bright regions inside granules, enclosed by the contours
in the lower left panel of Fig.~\ref{fig:comp_deconobs}.

\begin{figure*}[t]
	\vspace*{-3mm}	
		\centering
\includegraphics[width=15.5cm]{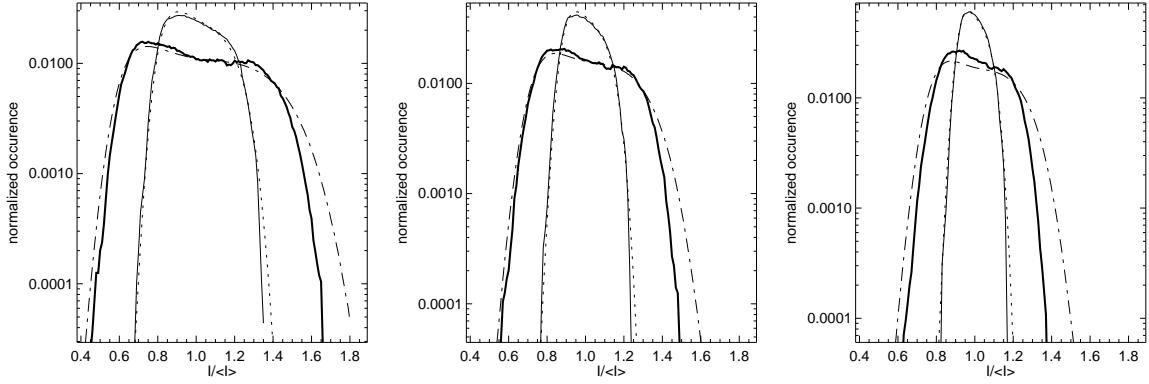}
	\vspace*{-3mm}	
\caption{Comparison of the intensity distribution in all synthetic and observed 
  disc-centre images for the blue (left), green (middle), and red continuum 
  channel (right). 
  Shown are the distributions for the 
  original observed images (dotted),  the deconvolved images (dot-dashed), 
  the degraded synthetic images (thin solid), and the original
  synthetic intensity maps (thick solid).} 
\label{fig:comp_hist}
\end{figure*}

\subsection{Degraded images compared to original observations}
\label{sec:comp_degsyn}

Now we reverse the procedure by artificially degrading the synthetic images by 
convolution with the non-ideal PSFs described in Paper~I. 
Using the best-fit PSFs with Voigt contributions, reduces the contrast values of 
the synthetic image to 13.0\,\%, 8.6\,\%, and 6.0\,\% (average over three snapshots) 
for the blue, green, and red continuum, respectively (see Table~\ref{tab:irmscomp}). 
The resulting differences between the average observational and degraded synthetic 
values are then 0.2, 0.3, and $-0.2$ 
percent points for blue, green, and red continuum, respectively. 
The above mentioned degradation was performed with a PSF extending with a size 
given by the computational box of the model, i.e., 15\,\farcs4\,$\times$\,15\,\farcs4. 
The far wings of the PSF are this way neglected. 
We therefore repeat the degradation with a PSF with a size of 110\arcsec\,$\times$\,110\arcsec, 
which is comparable to the FOV of the observations. 
The synthetic image is repeated periodically  to fill the FOV. 
This procedure is warranted as periodic boundaries were used during the simulation run. 
The large synthetic images are now degraded with the full non-ideal PSF, including the 
far wings. 
At disc-centre, we find mean contrast values of 11.3\,\%, 7.4\,\%, and 5.1\,\%
for the blue, green, and red continuum, respectively (see Table~\ref{tab:irmscomp}). 
This corresponds to an effective reduction of the contrast ranging from 0.87 to 0.85 of the 
value for the small FOV. 
On the other hand, uncertainties in the parameters for the model of the non-ideal PSF contributions 
have a tendency to increase the contrast again towards the value found for the small FOV
(see Sect.~\ref{sec:adduncert}).  
We conclude that the observed and synthetic snapshots agree well in terms of intensity 
contrast when considering the uncertainties inherent to the applied PSF. 

Figure~\ref{fig:comp_degradsim} shows examples for an original observed and a synthetic 
image in the green continuum channel. The intensity distributions of the original observed and 
the degraded synthetic image agree well. 
The distributions of individual images and subfields, however, vary to some degree
in accordance to the variation in intensity contrast.  
Some cases show a narrower distribution, some exhibit a number of particularly bright pixels 
(as in Fig.~\ref{fig:comp_deconobs} for the deconvolved image), while the secondary 
component of brighter than average pixels (see see Sect.~\ref{sec:comp_deconobs}) 
is sometimes only noticeable as a distribution asymmetry. 
The distributions are generally close to those found for the synthetic images. 
See Fig.~\ref{fig:comp_hist} for a comparison of simulation and observation 
that includes all available disc-centre images.

\begin{figure}[h]
	\vspace*{-5mm}
	 \includegraphics{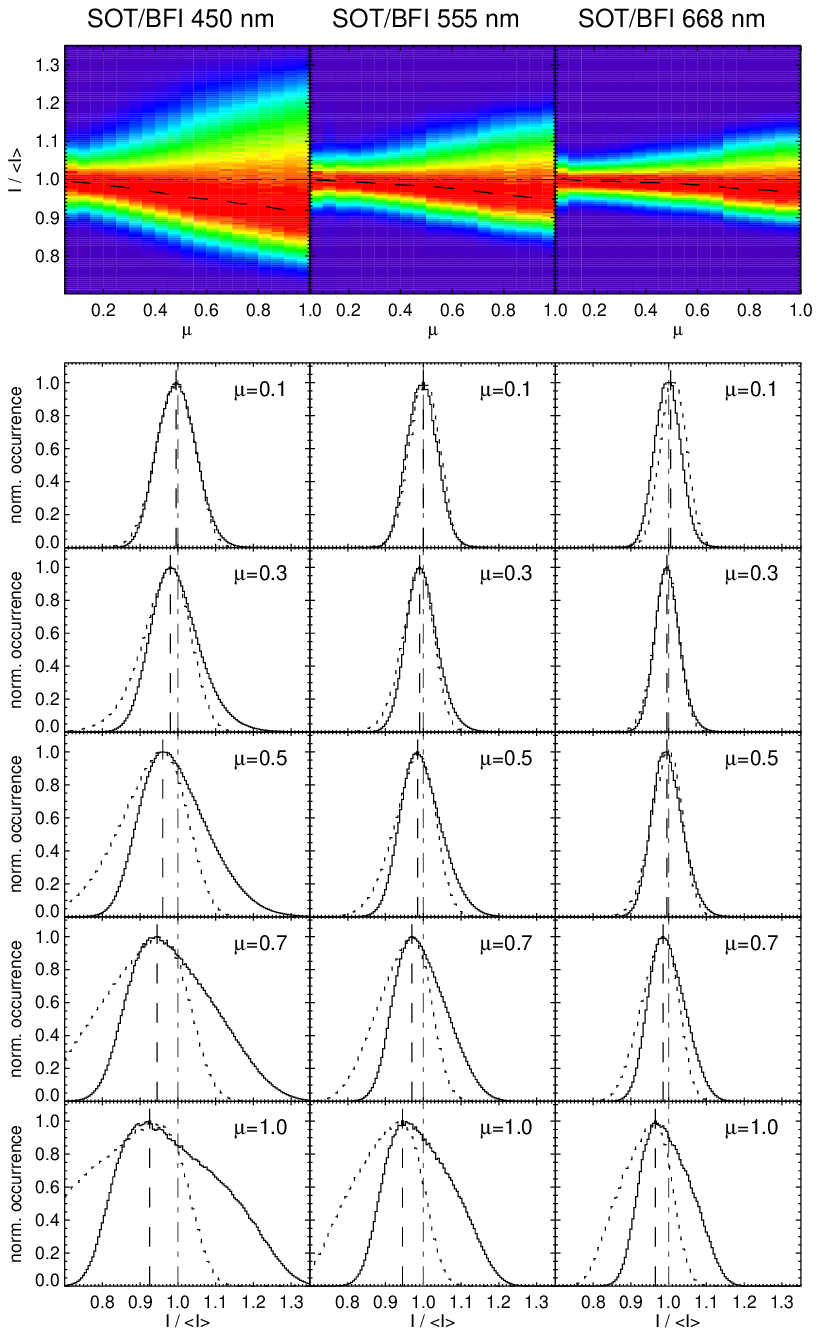}  
	 \vspace*{-3mm}
 	 \caption{Intensity histograms for the blue (left column), green (middle column),
   and red continuum (right column) as function of disc-position $\mu$.  
   \textit{Top row:} Histograms for all $\mu$ with the relative occurrence as 
   color/grey-shades.  
   \textit{Rows below:} Histograms for $\mu = 0.1$, 0.3, 0.5, 0.7, and 1.0, respectively.
	The lines mark $I/\langle{}I\rangle = 1$ (dot-dashed), the histogram peaks (dashed), 
	and -- in the lower rows -- the histograms mirrored at 
	the histogram peaks (dotted). 
	The latter reveals the asymmetry of the distribution.}
   \label{fig:sothist}
\end{figure}

\section{Centre-to-limb variation of the continuum intensity distribution}
\label{sec:intdis}

In this section, the properties of the observed and simulated intensity 
distribution and contrast are investigated as function of heliocentric position. 

\subsection{Observed intensity histograms.}
\label{sec:inthist}

The intensity distribution of the original (reduced) SOT images is illustrated 
in Fig.~\ref{fig:sothist} for the three continuum wavelengths. 
The pixels of all images are sorted by disc position $\mu$ into bins of size 
$\Delta \mu = 0.05$. 
The upper row of the figure shows the histograms for all bins, whereas the rows below 
show histograms for selected heliocentric positions. 
The intensity distribution becomes narrower with increasing wavelength 
and with decreasing $\mu$. 
The histograms are relatively symmetric close to the limb but exhibit a growing 
asymmetry towards disc-centre. 
The asymmetry is most pronounced at $\mu = 1.0$. 
Generally, there seems to be an increasing number of pixels with intensities  
brighter than average, which cause the distribution to be broader at disc-centre.  
Consequently, the peaks of the histograms move from around $I / <I> \approx 1.0$ 
at the limb to smaller values at disc-centre. 

Removing the PSF (see Sect.~\ref{sec:decon}) from the images has a dramatic 
effect on the resulting intensity distribution.   
It broadens, covering a much larger range in $I/\langle{}I\rangle$, 
and exhibits a subtle secondary peak for brighter than average pixels.  
More details are discussed in comparison to synthetic images in 
Sect.~\ref{sec:compwsim} and Fig.~\ref{fig:comp_deconobs}. 

\begin{figure*}[t]
  	\centering
   \includegraphics{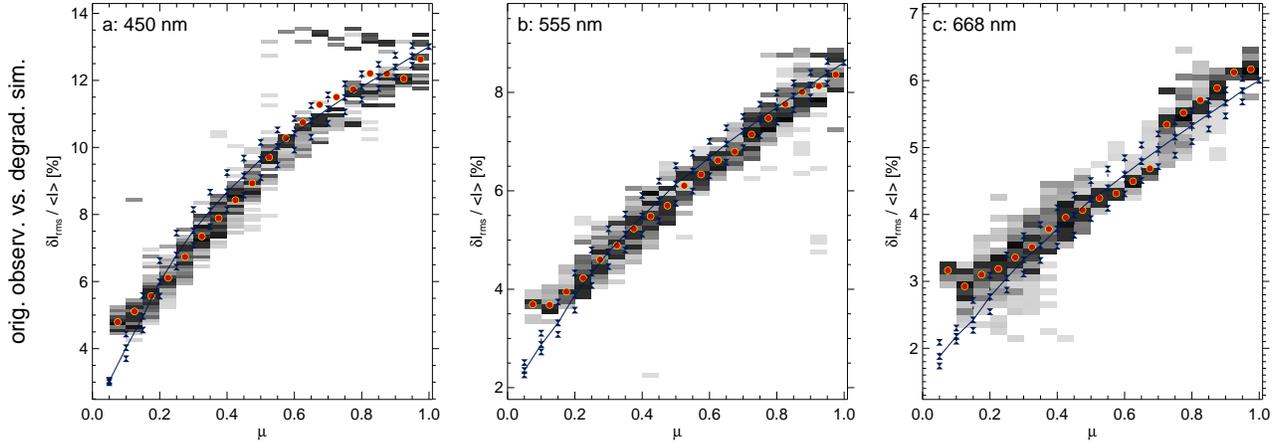}
   \caption{Centre-to-limb variation of the intensity contrast for 
   the blue (left column), green (middle), and red continuum (right). 
   The contrast values for the SOT wide-band filtergrams are 
   derived from regions of equal $\mu$  (bin size $\Delta \mu =0.05$) for all 
   original images. 
   The individual histograms are combined to a set of 
   $\mu$-dependent histograms (``density map''), which are shown 
   as grey scales. 
   The maxima of the histograms are marked with solid circles for each $\mu$ bin.  
   For comparison, the contrast values for the 3 Ch0 snapshots are shown
   as double triangles, while the solid line represents the mean centre-to-limb
   variation of model Ch0.}
   \label{fig:clv1}
\end{figure*}
\begin{figure*}[t]
  	\centering
   \includegraphics{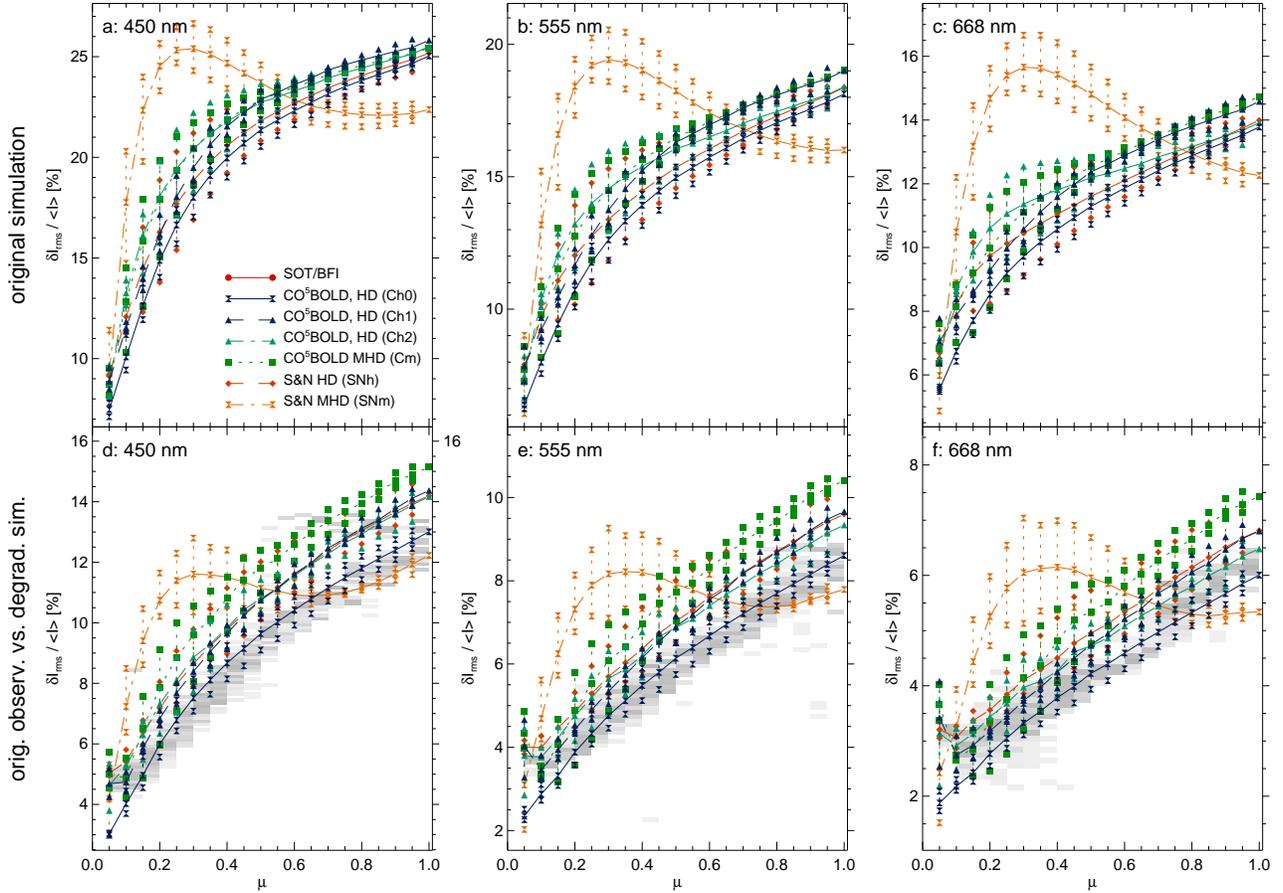}
   \caption{Centre-to-limb variation of the intensity contrast for all models: 
   \textit{top:} original synthetic images, \textit{bottom:} 
   after degradation with the non-ideal PSFs.  
   The columns show the results for the blue (left), 
   green (middle), and red continuum band (right), respectively. 
   The contrast values for the original SOT wide-band filtergrams are 
   represented as gray-shades for reference (cf. Fig.~\ref{fig:clv1}) in 
   the lower row.  The snapshots of each model are represented by symbols,
   while the lines show the mean centre-to-limb variation.
   Please note that only the actual simulation box sizes are
   considered for the degradation of the synthetic images shown here.
   Expanding it to a FOV comparable to the SOT observations would
   lower the individual contrast values but would give practical
   problems for low $\mu$ (see Sect.~\ref{sec:adduncert}).}
   \label{fig:clv2}
\end{figure*}

\subsection{Observed intensity contrast}
\label{sec:contrast}

\begin{table}[pb]
  \caption[h]{Intensity contrast \mbox{$\delta I_\mathrm{rms}$} in percent points 
  for the original observations in the three considered BFI continuum bands.}
    \label{tab:contrast}
  \centering
  \begin{tabular}[b]{c|cccccc}
  \hline
  \hline
  $\delta I_\mathrm{rms}^{\mathrm{\,a}}$&\multicolumn{6}{c}{$\lambda$}\\
  $\mu$$^{\mathrm{b}}$&\multicolumn{2}{c}{450.45\,nm}&\multicolumn{2}{c}{555.00\,nm}&\multicolumn{2}{c}{668.40\,nm}\\
  \hline
  \hline
 0.05 -  0.10&    4.8&(    5.0)&    3.7&(    3.8)&    3.2&(    3.2)\\
 0.10 -  0.15&    5.1&(    5.1)&    3.7&(    3.6)&    2.9&(    3.2)\\
 0.15 -  0.20&    5.5&(    5.4)&    4.0&(    4.0)&    3.1&(    3.2)\\
 0.20 -  0.25&    6.1&(    6.0)&    4.0&(    4.0)&    3.2&(    3.2)\\
 0.25 -  0.30&    6.7&(    6.6)&    4.3&(    4.1)&    3.4&(    3.2)\\
 0.30 -  0.35&    7.3&(    7.4)&    4.6&(    4.6)&    3.5&(    3.6)\\
 0.35 -  0.40&    7.9&(    8.0)&    4.9&(    4.8)&    3.8&(    3.5)\\
 0.40 -  0.45&    8.4&(    8.2)&    5.2&(    5.1)&    4.0&(    3.8)\\
 0.45 -  0.50&    8.9&(    8.9)&    5.4&(    5.4)&    4.1&(    4.0)\\
 0.50 -  0.55&    9.5&(    9.7)&    6.1&(    6.1)&    4.2&(    4.4)\\
 0.55 -  0.60&   10.0&(   10.4)&    6.2&(    6.4)&    4.3&(    4.3)\\
 0.60 -  0.65&   10.4&(    9.9)&    6.4&(    6.6)&    4.5&(    4.5)\\
 0.65 -  0.70&   10.8&(   10.7)&    6.6&(    6.4)&    4.7&(    4.7)\\
 0.70 -  0.75&   11.1&(   11.0)&    7.1&(    7.1)&    5.3&(    5.4)\\
 0.75 -  0.80&   11.3&(   11.4)&    7.5&(    7.5)&    5.5&(    5.6)\\
 0.80 -  0.85&   11.9&(   11.6)&    7.7&(    7.4)&    5.7&(    5.6)\\
 0.85 -  0.90&   12.1&(   12.2)&    7.9&(    7.8)&    5.9&(    5.8)\\
 0.90 -  0.95&   12.4&(   12.5)&    7.9&(    7.8)&    6.1&(    6.2)\\
 0.95 -  1.00&   12.8&(   13.1)&    8.3&(    8.4)&    6.2&(    6.2)\\
\hline
\hline
  \end{tabular}
  \begin{list}{}{}
  \item[$^{\mathrm{a}}$] 
  The average contrast in each bin is given next to the peak of the corresponding 
  contrast histograms (in parentheses, compare to Fig.~\ref{fig:clv1}). 
  \item[$^{\mathrm{b}}$] 
   heliocentric positions~$\mu$; The $\mu$ bins ($\Delta \mu =0.05$) are the same as in Fig.~\ref{fig:clv1}. 
  \end{list}
\end{table}

\begin{table}[pb]
  \caption[h]{Average contrast $\delta I_\mathrm{rms}$  and standard deviation of 
  the original and the deconvolved disc-centre observations in comparison with the 
  synthetic images. }
  \label{tab:irmscomp}
  \centering
  \begin{tabular}[b]{l|rlrlrl}
  \hline
  \hline
  $\delta I_\mathrm{rms}$ &\multicolumn{6}{c}{$\lambda$}\\
  $[\%]$&\multicolumn{2}{c}{450.45\,nm}&\multicolumn{2}{c}{555.00\,nm}&\multicolumn{2}{c}{668.40\,nm}\\  
  \hline
  \multicolumn{7}{c}{observation}\\
  \hline
  original     & 12.8&$\pm$ 0.5&  8.3&$\pm$ 0.4&  6.2&$\pm$ 0.2\\
  ideal PSF    & 18.0&$\pm$ 0.8& 12.3&$\pm$ 0.8&  9.9&$\pm$ 0.4\\
  full PSF     & 26.7&$\pm$ 1.3& 19.4&$\pm$ 1.4& 16.6&$\pm$ 0.7\\
  \hline
  \multicolumn{7}{c}{simulation Ch0 (Steffen 2007)$^{\mathrm{a}}$}\\
  \hline
  original     & 25.0&$\pm$ 0.1&  18.1&$\pm$ 0.1&13.8    &$\pm$ 0.1   \\   
  ideal PSF    & 16.7&$\pm$ 0.3&  11.4&$\pm$ 0.2&8.2     &$\pm$ 0.2   \\   
  non-ideal PSF (1)&
                 13.0&$\pm$ 0.3&   8.6&$\pm$ 0.2&6.0     &$\pm$ 0.2   \\   
  non-ideal PSF (2)& 
  		         11.3&$\pm$ 0.2&   7.4&$\pm$ 0.2&5.1     &$\pm$ 0.2   \\   
  \hline
  \multicolumn{7}{c}{other simulations (original)$^{\mathrm{b}}$}\\
  \hline
   Ch1         & 25.8&$\pm$ 0.5&  19.0&$\pm$ 0.5&14.6  &$\pm$ 0.5 \\   
   Ch2         & 25.5&$\pm$ 0.5&  18.4&$\pm$ 0.5&13.9  &$\pm$ 0.5 \\   
   Cm          & 25.4&$\pm$ 0.1&  19.0&$\pm$ 0.5&14.7  &$\pm$ 0.1 \\
   SNh         & 25.2&$\pm$ 0.6&  18.4&$\pm$ 0.4&14.0  &$\pm$ 0.3 \\
   SNm         & 22.4&$\pm$ 0.5&  16.0&$\pm$ 0.3&12.3  &$\pm$ 0.2 \\
  \hline 
  \hline
\end{tabular}
  \begin{list}{}{}
  \item[$^{\mathrm{a}}$] 
  The effect of the deconvolution with an ideal and a detailed PSF is specified for 
  the model by Steffen (Ch0). 
  The extent of the detailed PSF corresponds to 
  (1)~only the size of the model and 
  (2)~a periodic repetition of the model to a size of 
  110$\arcsec\,\times\,$110\arcsec, which is comparable to the FOV of the 
  observations. 
  \item[$^{\mathrm{b}}$] 
  For comparison, the original, i.e., non-degraded contrast values for sets of 
  \mbox{CO$^5$BOLD models}~(Ch1, Ch2, and Cm) and simulations by Stein~\& 
  Nordlund (SNh and SNm) are given. 
  \end{list}
\end{table}

\paragraph{Definition.} 
The intensity contrast is defined as
\begin{equation} 
 \delta I_\mathrm{rms} \equiv 
  \frac{\sqrt{ \frac{1}{N} \sum_{x, y} \left( I\,(x,y) - 
  \langle I\rangle_{x,y}\,\right)^2}
  }{\langle I\rangle_{x,y}} 
  \enspace,
\end{equation}
where $I\,(x,y)$ are the intensity values of the individual pixels with 
the average $\langle I\rangle_{x,y}$ over all $N$ pixels. 

\paragraph{Contrast of reduced images.} 
The contrast values are listed for the three considered BFI continuum bands 
and different heliocentric positions $\mu$ in Table~\ref{tab:contrast}. 
The average contrast values for the SOT filtergrams are derived from regions of equal 
$\mu$  with a bin size of $\Delta \mu = 0.05$ for all images.
At disc-centre, the images are within the same $\mu$-bin, whereas image segments are 
sorted into the bins closer to the limb. 
The complete images at disc-centre (15 for each wavelength), 
produce the following average contrasts and standard deviations:  
$(12.8\,\pm\,0.5)\,\%$, $(8.3\,\pm\,0.4)\,\%$, and $(6.2\,\pm\,0.2)\,\%$
for blue, green, and red continuum, respectively. 
The intensity contrast strongly decreases towards the longer wavelengths. 
The same behavior is found for all~$\mu$. 
For each wavelength, the contrast also decreases from centre to limb, i.e. 
with decreasing $\mu$, which can also be seen from the width of the histograms 
discussed in Sect.~\ref{sec:inthist}. 
This centre-to-limb variation of the contrast is shown in Fig.~\ref{fig:clv1}, which 
can be interpreted as grey-shaded histograms. 
The histograms for the individual bins exhibit a maximum close to the corresponding 
average contrast (solid circles, cf. Table~\ref{tab:contrast}). 
The distribution, however, cannot be described by means of a, e.g., simple Gaussian.  
The contrast distribution for the blue continuum even shows a small number 
of cases with enhanced contrast, which manifest as a secondary ridge in 
\mbox{Fig.~\ref{fig:clv1}a}. 
This is most likely connected to the influence of magnetic fields,
e.g., in the form of network bright points, which are unavoidable 
for the large size of the used FOV, even though enhanced network regions 
were explicitly excluded from the sample (see Sect.~\ref{sec:compwsim}). 
Small errors in the determination of the limb  and also the limited spatial 
resolution of SOT introduce uncertainties for the determination of the contrast 
very close to the limb (see Sect.~\ref{sec:disclimbobs}). 
Data for $\mu < 0.07$ should thus be interpreted with caution. 

\paragraph{Contrast after PSF deconvolution.} 
Removing the instrumental influence by deconvolution with the PSF has a 
dramatic effect on the contrast. 
The results are summarised in Table~\ref{tab:irmscomp}. 
The contrast of blue continuum images at disc-centre is increased by 
5.2~percent points after deconvolution with the diffraction-limited ideal PSF 
and another 8.7\,points, when the full non-ideal PSF is used. 
In total, the original contrast is more than doubled to $(26.7\,\pm\,1.3)\,\%$. 
The green and red channel show similarly large effects. 
The contrast increases from $(8.3\,\pm\,0.4)\,\%$ to 
$(19.4\,\pm\,1.4)\,\%$ in the green  
and from $(6.2\,\pm\,0.2)\,\%$ to $16.6\,\pm\,0.7)\,\%$ in the red channel.

\paragraph{Error estimate.} 
It is not trivial to give a reliable error estimate for the empirical contrast values. 
Next to systematic errors due to instrumental properties (PSF, straylight, see 
Sect.~\ref{sec:adduncert}), there are 
statistical errors arising from the selection of the pixel ensembles. 
The contrast values given in Table~\ref{tab:contrast} are derived from pixels with 
similar heliocentric position $\mu$, effectively forming segments of circles in the 
images. 
At disc-centre, the change in $\mu$ over one BFI image is so small  that the 
complete images instead of segments are used.
The rms values in Table~\ref{tab:irmscomp} thus represent the variation among whole 
images. 
The smaller the $\mu$ bin-size, the smaller the number of the included pixels over 
which intensity mean and variation are computed.  
This can lead to a small increase of the scatter in contrast at small $\mu$, 
while the results for positions closer to the disc-centre are less susceptible.  
Another way to determine a statistical error is to compare the variation of the 
contrast in square subfields.  
This error increases slightly with decreasing subfield size. 
While the same rms variations than in Table~\ref{tab:irmscomp} are found for large 
field sizes, the spread increases slightly to 0.7\,\%, 0.6\,\%, and 0.3\,\% 
for blue, green, and red continuum, resp., when dividing the images into fields  
of only 10\,\arcsec\,$\times$\,10\,\arcsec. 

For the construction of the optimal filter used for deconvolution, assumptions 
have to be made for the signal model. 
As already stated in Sect.~\ref{sec:decon}, the two different signal models 
considered here result in contrasts that differ by only $\sim 0.2$ percent points. 
The deconvolution procedure seems to be rather robust. 
We degrade the deconvolved images again with the same PSF in exactly the way as it 
is done for the simulations in Sect.~\ref{sec:compwsim}. 
The original observed images and the back-transformed images have contrast values 
that differ by much less than 0.1 percent points.

\subsection{Synthetic intensity contrast}
\label{sec:synic}

For $\mu = 1$, the three selected snapshots of the model Ch0 give an average 
contrast of 25.0\,\%, 18.1\,\%, and 13.8\,\%, at the central wavelengths of the blue, 
green, and red continuum channels of the BFI, respectively. 
The variation among the individual snapshots is for all wavelengths of the order of 
0.1~percent point. 

Generally, there is a spread of a few tenths of a percent point among individual 
snapshots.  
All non-magnetic \mbox{CO$^5$BOLD} models and also SNh produce very similar 
contrasts. 
Obviously, the model Ch0 analysed in more detail in Sect.~\ref{sec:compwsim} 
is no exception but representative for state-of-the-art 
granulation simulations.  
Also the weak field model Cm shows no systematic difference with respect to the 
non-magnetic models. 
In contrast, the model with the high field strength (SNm) exhibits systematically 
lower contrasts with respect to SNh at disc-centre. 
The values nevertheless remain roughly of the same order as for the other models 
but become significantly larger at smaller $\mu$ (see Fig.~\ref{fig:clv2}). 

The contrast of all non-magnetic and weak field models decreases monotonically with decreasing $\mu$. 
The difference between the individual simulations can essentially be described as a small contrast offset. 
An interesting exception is the higher field strength model SNm. Starting with a lower value 
than SNh at disc-centre, the contrast grows with decreasing $\mu$ until a peak 
is reached at $\mu = 0.3$. For $\mu < 0.8$, SNm exceeds the contrast of SNh. 
The blue continuum band of SOT in Fig.~\ref{fig:clv1}a indeed shows a secondary group 
of image segments with a contrast excess, which starts to divert from the main distribution 
already at disc-centre.   
It is not obvious at which $\mu$ this subset of SOT observations reaches maximum contrast
but it could well be for smaller $\mu$ than seen in the model SNm. 
The latter represents a plage region with $\langle|B|\rangle = 250$\,G, whereas 
the effect is not clearly discernible for the weak-field model Cm. 
We therefore argue that the subset of SOT data with contrast excess might contain 
a small number of magnetic field structures with smaller field strengths.

The deviation of the contrast is connected to the appearance of solar faculae 
close to the limb. 
Faculae can be understood as a line-of-sight effect caused by magnetic elements in the 
intergranular lanes. 
This topic has been discussed extensively in recent studies by  
\citet{2004ApJ...607L..59K}, \citet{2004ApJ...610L.137C}, \citet{2005A&A...430..691S}, 
\citet{2006ApJ...646.1405D}, and \citet{2007ApJ...661.1272B} 
but also by \citet{2005A&A...438.1059H}, \citet{2003A&A...397.1075S}, \citet{1999SoPh..189...57S}, and many more.
See \citet{2002A&A...388.1036O} for an empirical investigation of 
the centre-to-limb variation of the intensity contrast for solar photospheric faculae. 

The additional high-contrast ridge seen in the blue is not visible for the other
two SOT channels in  Fig.~\ref{fig:clv1}.
This might partly be a selection effect as we explicitly selected quiet Sun images, 
avoiding enhanced magnetic field whenever possible. 
The images were taken during solar minimum so that the Sun was very quiet 
and not many pronounced active regions were present.  
Still one has to consider weak magnetic fields on small spatial scales in the quiet 
Sun. 

The small contrast differences among the non-magnetic models are partly systematic effects due 
to differences in the radiative transfer schemes and in the opacity tables, e.g., PHOENIX/OPAL 
opacities \citep{hauschildt97,opal} for Ch1. 
On the other hand, already the random selection of three snapshots influences the mean contrast
by the order of a few tenths of a percent point. 
A larger set of 63 images for Ch2, for instance, produces contrasts in the blue continuum 
between 23.7\,\% and 26.7\,\% with a mean of 25.2\,\% and an rms variation of 0.7\,\%. 
While the mean is reduced by 0.3 percent points for the larger sample, the rms variation increases 
from 0.5\,\% to 0.7\,\% (cf. Table~\ref{tab:irmscomp}).  
A determination of the mean granulation contrast with a precision down to the order of a few 
0.1\,\% obviously requires a statistically significant sample of images rather than a single snapshot. 

\section{Spatial power spectral density}
\label{sec:spatpower}

\subsection{Observed power spectral density}

\begin{figure*}[t]
   \sidecaption	
   \includegraphics[width=12cm]{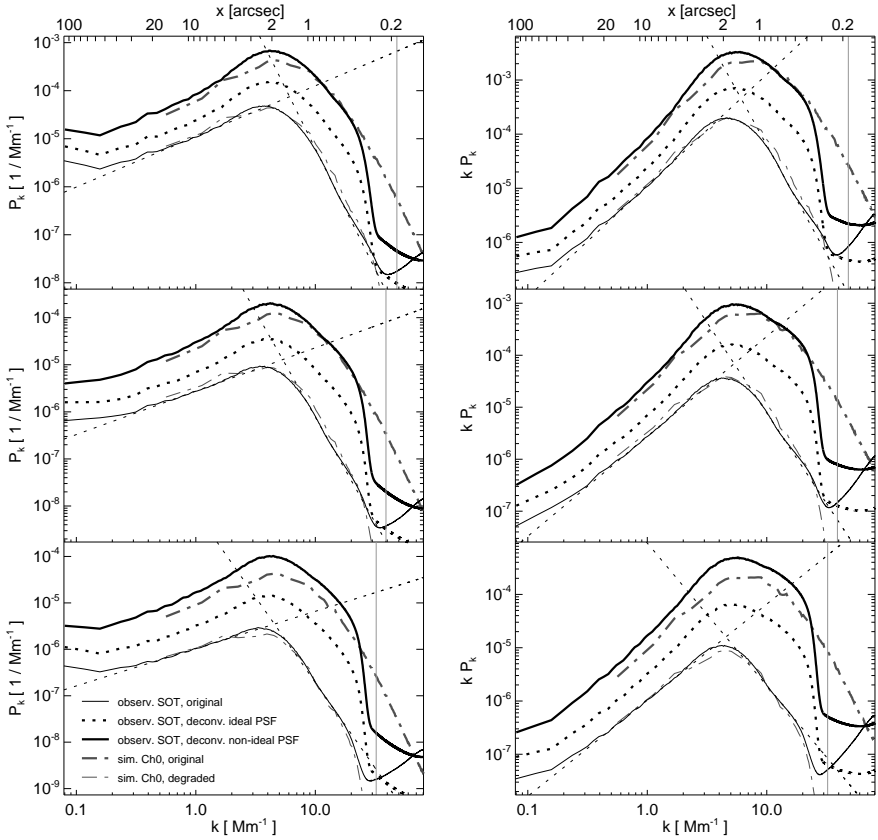}
   \caption{
   Power spectral density $P$ (left column) and $k P_k$ (right column) for 
   $I/\langle{}I\rangle$ as 
   functions of spatial wavenumber $k$ for the three BFI continuum wavelengths at  
   disc-centre: blue (top row), green (middle), and red continuum (bottom). 
   The observational data (thin solid) is compared to the 
   the degraded synthetic data (thin dot-dashed), while the 
   original synthetic images (thick dot-dashed) are checked against 
   the observations after deconvolution with the ideal PSFs (thick dotted) and 
   the non-ideal PSFs (thick solid).  
   The slopes of the power spectral density can be approximated with power-laws of 
   the form $P_k \propto k^{s_P}$. 
   The exponents are \mbox{$s_P \sim 1.0$} for small $k$ and between $\sim -4.0$ 
   and $\sim - 5.0$ for large $k$ (dotted lines). 
   For ${k P_k}$, the exponents are correspondingly $s_{kP} = s_P + 1$
   (see text for details). 
   The dotted lines represent the power law fits to the observational data.  
   The solid vertical lines indicate the wavenumber corresponding to the 
   diffraction limit of the telescope.}  
   \label{fig:power}
\end{figure*}

The power spectral density $P_k$ (PSD or short "power spectrum") is calculated  by 
integrating the amplitude square $H^2$ of the two-dimensional Fourier transform of 
an image over rings of constant wavenumber $k$. 
We use wavenumbers defined as 

\begin{equation} 
	k = 
	\sqrt{k_x^2 + k_y^2} = 
	2\,\pi\,\sqrt{{L_x}^{-2}\,+\,{L_y}^{-2}}\enspace,
\end{equation}
where $L_x$ and $L_y$ are the spatial wavelengths in $x$- and $y$-direction.  
Please note that here the {\em one-sided} PSD is used, i.e. the Fourier 
amplitudes $H$
for both the negative {\em and} the positive wavenumbers are considered   
\citep[see, e.g.,][]{1992nrfa.book.....P}. 
As the Parseval theorem is fulfilled, the total power for the intensity 
$I / \langle I \rangle$ is related to the contrast of the original image: 

\begin{equation} 
\label{eq:parseval}
\int P_k\, dk' = 
\left(\delta I_\mathrm{rms}\,\right)^2
\enspace.
\end{equation}
Power spectra are calculated for granulation images at disc-centre
for all three considered SOT continuum bands. 
The azimuthally integrated power spectra are plotted versus $k$ in the left column 
of Fig.~\ref{fig:power}. 
The curves show power maxima at a spatial scale of 
$\sim 1.7$\,Mm, 1.8\,Mm, and 1.9\,Mm for the blue, green, and 
red continuum, respectively. 
That scale is somewhat larger than the most frequent granule diameter, which is 
typically of the order $\sim 1.4$\,Mm in the SOT images analysed here.
The latter is in very good agreement with earlier 
results such as, e.g., with the 1\,\farcs9 reported by \citet{1977SoPh...54..319B}. 
The spectra also show signs of enhanced power density at spatial scales $> 20$\,Mm 
that correspond to super-granulation. 
This is in line with \citet{2008A&A...479L..17R}, who derive the kinetic energy 
density spectrum by granule tracking.  
Except for these additional contributions, 
the slopes can roughly be approximated with power laws of the form $P_k \propto k^{s_P}$. 
The exponents $s_P$ for small $k < 2.0$\,Mm$^{-1}$ are of the order of 
$1.0$, i.e. the power increases essentially linearly with wavenumber. 
Fits of the observational data produce exponents that decrease slightly with wavelength 
from 1.0 for the blue, to 0.9 for the green, and 0.8 for the red channel. 
Power law fits are also possible for large wavenumbers although the exponents change with 
$k$ and thus depend on the considered wavenumber range. 
In the range from 9\,Mm$^{-1}$ to 18\,Mm$^{-1}$, we find exponents of 
roughly $-5.0$, $-4.5$, and $-3.7$ for blue, green, and red continuum, respectively.
The slopes get steeper at even higher wavenumbers before the power spectral density 
is dominated by noise and increases again. 
The contributions to the power spectral density at these wavenumbers are, however, 
of minor importance only. 

The right column of Fig.~\ref{fig:power} shows the quantity $k P_k$. 
According to Eq.~(\ref{eq:parseval}), the area below the curve is equal to the intensity 
contrast squared. 
The distributions have clear peaks corresponding to those in the power spectra in the 
left column but at somewhat smaller spatial scales: 1.4\,Mm for blue and red continuum 
and at 1.5\,Mm for the green channel. 
These scales coincide with the most frequent granule diameter of $\sim 1.4$\,Mm (see above).   
The slopes can again be approximated with power laws of the form 
$k P_k \propto k^{s_{Pk}}$. 
The exponents $s_{Pk}$ should correspond to  $s_{kP} = s_P + 1$ and are indeed found 
to be in the range between $-4.0$ and $-2.7$.

\subsection{Synthetic power spectral density}
Not only the intensity distribution but also the power spectral density of 
model Ch0 is in line with other simulations as can be seen, e.g., 
from Fig.~3 by \citet[][]{2000SoPh..192...91S}. 
It illustrates that model Ch0 is representative of state-of-art numerical 
simulations of solar granulation. 

\subsection{Comparison of observed and synthetic PSD}

Figure~\ref{fig:power} demonstrates that the power spectral density of the degraded 
synthetic images matches well the observational data described 
in Sect.~\ref{sec:spatpower} for all three continuum bands.  
The power law fits for the degraded synthetic data produce exponents that 
agree very closely with the empirical exponents. 
The slope for larger wavenumbers in the red continuum, however, is better 
represented with an exponent of $s_P \approx -4.0$, which is slightly smaller 
than the observational analogue. 
At these wavenumbers, the instrumental image degradation  
becomes significant and, by suppressing power at these scales, influences 
the slopes. 
The resulting steep decrease in power spectral density is also present for the 
degraded synthetic images. 

The power spectral density for the original synthetic images is generally higher 
corresponding to a higher intensity contrast (see Eq.~(\ref{eq:parseval})). 
The power law exponents are determined as 1.0 for small wavenumber
and $\sim -4.5$ at large wavenumbers for all three wavelengths and 
thus still agree with the observations.  
The slight wavelength dependence found for the observed and the degraded data is not 
present or at least less pronounced in the original synthetic intensity maps. 
Furthermore, the power spectral density of the latter continues to decrease 
further with wavenumber well into the domain  where the observational signal 
(and also the degraded synthetic one) is drowned in noise (see 
Fig.~\ref{fig:power}), i.e., $k > 40\,\mathrm{Mm}^{-1}$. 
Due to that steep slope, the integral of the PSD is changed only slightly when 
going towards higher wavenumbers. 
According to the normalisation in Eq.~(\ref{eq:parseval}), a small change of the integral 
corresponds to a small change in intensity contrast. 
The contribution to the intensity contrast  is therefore  
very small at high wavenumbers but large for granulation scales corresponding to smaller~$k$.

\begin{figure}[t]
\centering
\includegraphics{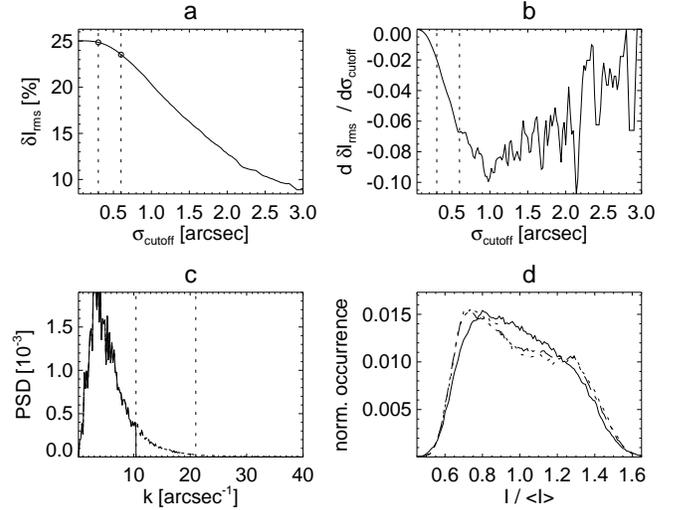}
\caption{Fourier-filtering of synthetic images at 450\,nm. 
Intensity contrast \mbox{$\delta I_\mathrm{rms}$} as function of the imposed 
spatial cutoff $\sigma_\mathrm{cutoff}$ (\textbf{a}) and its 
derivative (\textbf{b}).
The power spectral density (\textbf{c})  and the 
intensity distributions (\textbf{d}) are plotted for the 
original images (dotted lines) and two 
selected cutoffs of 0\,\farcs3 (dot-dashed) and 0\,\farcs6 (solid), which 
are marked as open circles and/or vertical dotted lines in panels~a-c.}
\label{fig:fftfilter}
\end{figure}

We illustrate this with an experiment in which we artificially filter away the 
small scales in synthetic images for the blue band. 
It is done by Fourier-transforming an image, eliminating all Fourier-components 
for wavenumbers larger than an imposed cutoff $k_\mathrm{cutoff}$, and back-transforming. 
The power spectral density of the resulting image is thus truncated at $k_\mathrm{cutoff}$
(see Fig.~\ref{fig:fftfilter}c). 
The intensity contrast of the filtered images is plotted in Fig.~\ref{fig:fftfilter}a 
as function of the spatial cutoff scale $\sigma_\mathrm{cutoff} = 2 \pi / k_\mathrm{cutoff}$. 
The contrast decreases monotonically with the cutoff. 
Towards the smallest scales, \mbox{$\delta I_\mathrm{rms}$} saturates to the value of 
unfiltered images (25.0\,\%). 
The asymptotic approach to the final value already starts at scales that are
well resolved in the numerical model (grid size $\sim 0\,\farcs039$). 
The change of the slope can be seen from derivative 
\mbox{$\delta I_\mathrm{rms}/\sigma_\mathrm{cutoff}$} 
in Fig.~\ref{fig:fftfilter}b. 
(The derivative at larger $\sigma_\mathrm{cutoff}$ is noisy due to the numerical 
treatment of the discrete Fourier filtering.) 
The biggest change of the contrast occurs at scales between 
$> 0\,\farcs5$ and $\sim 2\,\farcs2$, 
which coincides with the ``granulation peak'' in the PSD. 
In contrast, the change towards the smallest scales decreases quickly. 
Removing the smallest scales has apparently only a small effect in terms of contrast. 
A cutoff of $\sigma_\mathrm{cutoff} = 0\,\farcs5$ still produces a contrast of 24.2\,\%, i.e., 
a reduction of only 0.8 percent points.  
The absence of small scales is more obvious when looking at the intensity histograms 
in Fig.~\ref{fig:fftfilter}d. 
The distributions for the unfiltered images and for 
$\sigma_\mathrm{cutoff} = 0\,\farcs3$ are very similar, whereas 
$\sigma_\mathrm{cutoff} = 0\,\farcs6$ already produces noticeable differences. 
The otherwise clearly visible double-peak transforms into a single-peaked asymmetric distribution. 
The reduced occurrence of the brightest and darkest pixels implies that they are connected 
to the smallest scales.  
The darkest pixels, which constitute the `major component of the 
intensity distribution (at $I < \langle{}I\rangle$, cf. Sect.~\ref{sec:comp_deconobs}
and Figs.~\ref{fig:sothist} and \ref{fig:comp_deconobs}),   
are found in the narrow intergranular lanes. 
In the case of SOT, the lanes seem to be sufficiently resolved. 
We therefore expect no change of the contrast numbers for deconvolved images 
from the use of telescopes with larger apertures.  
See, e.g., \citet{2001ChA&A..25..439R} for an analysis of 
the granulation contrast as function of the telescope aperture. 

Finally, we compare the PSD derived from the deconvolved observations with 
those for the original synthetic images, which are all plotted as 
thick lines in Fig.~\ref{fig:power}. 
The dotted and solid lines represent the PSD for deconvolution with the ideal 
and the non-ideal PSFs, respectively. 
The original simulations are marked with the dot-dashed lines. 
While deconvolution with the ideal PSF already produces a significant increase in 
power througout the observationally accessible wavenumber range, 
the non-ideal PSFs bring the observations very close to the simulation results. 
The PSDs agree well for the blue and green continuum. 
For the red continuum, however, we remain with a small discrepancy, which is 
consistent with the differences in terms of intensity contrast 
(see Table.~\ref{tab:irmscomp}). 
The sharp decrease of the deconvolved PSDs close to the theoretical cutoff 
wavenumber (vertical lines in Fig.~\ref{fig:fftfilter}) results in a significant 
power deficit with respect to the simulations. 
This is simply due to the fact that there is no further information content in 
the images at the corresponding spatial scales due to the limited spatial resolution. 
Consequently, the power contributions at wavenumber close to the spatial cutoff 
are lost. 

\section{Discussion}
\label{sec:discus}

\subsection{Summary.}

Both simulations and observations are found to have a considerable variation 
in contrast. 
It is mostly a statistical variation between different snapshots / images. 
And still there are systematic, physical sources to be considered. 
\citet[][see also \citeauthor{1990ARA&A..28..263S} 1990]{1989ApJ...336..475T} have 
shown that removing the intensity modulation due to f- and p-mode oscillations in 
the data taken with SOUP at the former Swedish Vacuum Solar Telescope reduces the 
rms intensity variations by 35\,\%. 
The relative contribution of oscillations to the contrast depends on the spatial 
resolution as oscillations are correlated with certain spatial scales of which 
some possibly cannot be resolved \citep[see, e.g.,][]{1990ARA&A..28..263S}.

We conclude that in view of the mutual variation ranges one cannot expect a much 
better match of synthetic and observational continuum intensity contrast than has 
been achieved here.

\subsection{Comparison with other studies}

Observed contrast values for quiet Sun granulation vary significantly
throughout the literature, indicating the large uncertainties arising
from often unknown properties of the employed instrument and seeing
effects.
See \citet{2000ApJ...538..940S} for an extensive survey of observed contrast 
values. 
Here we discuss some of the more recent studies that report
granulation contrast measurements.

\citet{2003A&A...408.1115K} determined a granulation contrast between 5 and 6\,\%
for the continuum around 500\,nm from observations with the
VTT. 
Our SOT data rather indicate a contrast value of $\sim 11$\,\% for
flat-fielded images at this wavelength.
Note that \citet{2003A&A...408.1115K} make no corrections for
instrumental and seeing effects.

Also using the VTT, \citet{1995A&AS..114..387D} found a contrast of
14.5\,\% \citep[cf.][]{2002A&A...395..279P} in speckle reconstructed
images at 550\,nm, which is significantly lower than the 19.4\,\% we
derived for the green continuum channel of SOT/BFI after deconvolution
with the non-ideal PSF.

Van Noort et al. (\citeyear{2005SoPh..228..191V}) report granulation contrast 
numbers for SST observations in the continuum at 436.4\,nm.
Their images are restored with the MOMFBD image restoration technique
and have maximum contrast values slightly below 11\%.
These values are significantly lower than the 12.8\% that we find for the
SOT/BFI blue continuum at slightly longer wavelength {\it before}
deconvolution (see Table~\ref{tab:irmscomp}).
The PSFs they derived from MOMFBD processing are based on
sub-images of size $\sim$5\arcsec$\times$5\arcsec and can therefore
not account for the far wings of the PSF.
No correction for scattered light is applied through MOMFBD
processing. 
As we will discuss later in this section, this is probably the main 
reason why their contrast values are relatively low.

\citet{2007ApJ...668..586U} compare speckle-reconstructed G-band images of the 
quiet Sun taken with the Dunn Solar Telescope (DST) with a synthetic image calculated 
from a snapshot of radiation magnetohydrodynamical simulation by 
\citet{2006ApJ...642.1246S} with an initially horizontal magnetic field with 
$\langle{}|B_0|\rangle = 30$\,G. 
They find a considerable discrepancy between the observed contrast of 14.1\,\% for 
their best reconstructed images, and the synthetic contrast of 21.5\,\%.
\citeauthor{2007ApJ...668..586U} also analyse a synthetic image from  
the same \citeauthor{2006ApJ...642.1246S} simulation with $\langle{}B_z\rangle = 250$\,G 
that we use here (see Sect.~\ref{sec:sim}).   
They determine that the contrast of this snapshot, which corresponds to a plage 
region, is only 16.3\% in the G-band. 
The fact that the effect of increasing the magnetic field leads to a
decrease in the contrast, leads them to speculate that the quiet Sun
might contain a larger amount of weak magnetic field than the 30\,G
assumed for the quiet Sun model.

Our comparison of synthetic contrast values in the continuum around
$\lambda = 450$\,nm in Sect.~\ref{sec:synic} also shows that a higher 
average magnetic field strength lowers the contrast (see Table~\ref{tab:irmscomp}) 
but to much smaller extent than found by \citeauthor{2007ApJ...668..586U} 
in the G-band.
This suggests that the presence of a large amount of spectral lines in
the G-band is an important factor for the value of the granulation
contrast as one compares magnetically active and quiet regions.

We analyse speckle-reconstructed images obtained with the Dutch
Open Telescope \citep[DOT,][]{2004A&A...413.1183R}, which we obtained from 
the DOT online data base.
We find contrast values ranging between 14.4\,\% to 18.6\,\%  
for the G-band and values in the range from 14.9\,\% to 19.8\,\%
for the DOT blue continuum at 432\,nm. 
The code used to reconstruct the DOT images is based on an IDL code first 
mentioned in \citet{1992A&A...264L..24D}.
While some of the DOT contrast values are close to the DST result from 
\citet{2007ApJ...668..586U}, others are clearly higher. 
\citeauthor{2007ApJ...668..586U} find values around 14\,\% for G-Band images 
obtained with the DST using the speckle code described 
in detail in \citet{2008SPIE.7019E..46W}. 
This code is different from the code used for the aforementioned DOT observations.
The code by \citeauthor{2008SPIE.7019E..46W} takes into account the usage of an 
Adaptice Optics (AO) system. 
The functions used for the Fourier amplitude calibration are modified 
for Adaptive Optics usage according to \citet{2007ApOpt..46.8015W}. 
See \citet{2008A&A...488..375W} for an analysis of the code's 
photometric accuracy in comparison to the Hinode satellite. 

The previous paragraphs illustrate that there is a considerable spread among the 
contrast values of reconstructed images due to inherent sources of uncertainty. 
In the case of speckle interferometry, uncertainties in the Fourier amplitude 
calibration due to an incorrect estimation of the Fried parameter and/or (if applicable) 
the AO correction performance -- and thus the application of the wrong calibration 
function -- can lead to a large scatter in contrasts for the reconstructed images.

In general,  instrumental straylight is not accounted for in neither the 
speckle nor the MOMFBD reconstruction process, and
needs to be treated separately. 
Therefore, the contrast values from
post-facto reconstructed images should at best be compared to 
those disc-centre images of SOT that have been deconvolved with the ideal PSF only 
(e.g., (18.0$\,\pm\,$0.8)\,\% for the blue continuum). 
But even then it is not guaranteed that the effect of seeing is completely removed 
while this source of uncertainty is of no concern for the space-bourne measurements 
with Hinode. 

The apparent contrast dilemma can be solved by reviewing the problematic 
treatment of non-ideal PSF contributions. 
The determination of the straylight level is very difficult, in particular 
for ground-based observations. 
Consequently, details of the straylight properties are often unknown. 
A frequent strategy to account for straylight is to simply assume 
a constant contribution $\alpha$, which is equally spread over the FOV.  
It corresponds to an intensity offset, which decreases the intensity contrast to 
$(1 - \alpha)$ of its original value. 
\citet{2007ApJ...668..586U} conclude that they would 
need a straylight level of $\alpha = 0.34$ in order to reproduce the contrast in their synthetic images. 
They consider this a factor 3 too high for the DST.  
We also determine $\alpha$ from the difference between SOT images deconvolved with 
just the ideal diffraction-limited PSF and the full non-ideal PSF (see Table~\ref{tab:irmscomp}). 
In case of the blue continuum at disc-centre, the contrast difference of 
8.5\,percent points translates into 
$\alpha = 0.33$. 
For the green and red continuum, we find 0.37 and 0.41, respectively. 
Although these values are very close to the one by \citeauthor{2007ApJ...668..586U}, 
they should not be interpreted as actual straylight levels. 
The eclipse and Mercury transit observations, on which the SOT PSF with non-ideal 
Voigt-function contributions is based on, show 
residual intensities of the order of a few percent points only (see
Paper~I). 
Obviously, the assumption of a constant straylight contribution $\alpha$ is 
misleading. 
A constant offset would not only influence the PSF wings as intended but
also the PSF core. 
For SOT, the straylight level was actually shown to be dependent on the 
mean count rate, i.e., the overall light level in the instrument, and with that on 
the heliocentric position of the FOV. 
We therefore argue that a more detailed straylight model is crucial for detailed 
comparions of the granulation contrast and the intensity distributions in general. 

\citet{2008A&A...484L..17D} obtained a continuum intensity map at
630\,nm wavelength with the spectro-polarimeter (SP) of SOT. 
For this single map, they report a contrast of 7.0\.\%, a value which
is in line with our results from SOT/BFI.
After detailed modelling of the SOT/SP optical system, they degrade
a synthetic image from a MHD simulation by \citet{2007A&A...465L..43V}
taking into account a detailed PSF. 
For the degraded synthetic image, they find a contrast of 8.5\%.
\citet{2008A&A...484L..17D} argue that a slight defocus of the SP
instrument can account for a further decrease of the synthetic
contrast of 1\%.
The remaining 0.5\% discrepancy is then attributed to straylight and
imperfections in the optical system, although one could also argue for
an intrinsic spread in the observed contrast values.
Our results point at a range of observed contrast values on the order of
a few tenths of a percent around 630\,nm. 

In their recent paper, \citet{2009ApJ...694.1364B} present the rms granulation 
contrast derived from detailed 3D radiative transfer for the 3D hydrodynamic 
model by \citet{2000A&A...359..729A}. 
Judging from their Fig.~9, the synthetic intensity maps have contrasts of 
roughly 24\,\%, 17\,\%, and 12\,\% for wavelengths corresponding to the blue, 
green and red channels of the BFI. 
These values are systematically lower than for the synthetic maps considered 
here (see Table~\ref{tab:irmscomp}). 
The difference, however, is well within the variation range of individual 
observed images and uncertainties discussed in Sect.~\ref{sec:adduncert}. 

\subsection{Additional sources of uncertainty}
\label{sec:adduncert}

\paragraph{Spectrum synthesis:}
In Linfor3D, pure local thermodynamic equilibrium is assumed, 
i.e. scattering is not taken into account. 
In their Fig.~8, \citet{2009ApJ...694.1364B} present the centre-to-limb 
variation of the continuum intensity at \mbox{$\lambda = 300$\,nm.} 
They demonstrate that accounting for the scattering contribution to the source 
function removes the discrepancy between the observed and synthetic intensity. 
The discrepancy, however, is only of the order of a few percent points for 
small $\mu$ and vanishes towards disc-centre. 
Scattering is even less important for the wavelengths considered here. 
We therefore argue that the neglect of scattering in Linfor3D has only small 
if not negligible  influence on our results.   

\paragraph{Point Spread Function.}
As the conditions may vary somewhat between different observations, 
the corresponding PSF also varies. 
Consequently, each image would require its very own specific PSF for deconvolution. 
In practice, it was only possible to determine PSFs for BFI/SOT that represent the 
best-fit cases for a large range in observational conditions (see Paper~I). 
They may be considered as representative of a typical observational situation. 
The analysis of the deconvolved images should therefore be based on a statistically 
significant set of images. 

Also the degradation of the synthetic images with the PSF bears sources of 
uncertainty.
The spatial dimensions of the simulations used here are smaller than for the 
observed images and the extent of the PSFs from Paper~I. 
The convolution of a synthetic map with a PSF therefore mainly takes into account 
the central parts of the PSF while the far wings are effectively neglected.  
As mentioned already in Sect.~\ref{sec:comp_degsyn}, one should consider a 
FOV that is comparable to the one of the observation. 
Periodical extension of the synthetic maps to a size of 
110\,\arcsec$\,\times\,$110\,\arcsec\  and degradation by a 
correspondingly larger PSF, 
lead to a reduced intensity contrast for the model by Steffen 
(see Table~\ref{tab:irmscomp}). 
A lower contrast of the synthetic maps seems actually to be in line with 
the difference between the contrasts of the deconvolved observations and the 
original synthetic maps.

The influence of a too small FOV can also be seen for the other models in 
Fig.~\ref{fig:clv2}. 
Except for the model with higher magnetic field strength, all simulations produce 
contrast values for the undegraded images that lie within their mutual variation ranges
(see right column in the figure). 
After degradation (middle column), however, a systematic shift can be seen. 
The \mbox{CO$^5$BOLD} models are stacked according to the size of their spatial extent: 
low contrasts for model by Steffen  to higher contrasts for the smallest model Cm. 
The systematic difference can be removed by consistently going towards a large FOV of 
the same size. 
Making a realistic extension of the FOV is only practically feasible for disc-centre 
images.
For regions towards the limb, however, the extension of the FOV would require to vary the
observing angle across the image. 
As there is no straightforward procedure to do this, we refrain from enlarging the PSF for 
the calculation of the centre-to-limb variation of the contrast.  

The uncertainty in the PSF wing contribution (as seen from the choice of the FOV) is 
counteracted by uncertainties in the PSF parameters. 
In the following, we illustrate the effect for the blue continuum images. 
So far we used the best-fit PSF as determined in Paper~I. 
The PSF for the blue BFI channel is determined by the Voigt parameters 
$\gamma = 0\,\farcs004$ and $\sigma = 0\,\farcs008$ (see Sect.~\ref{sec:psf}). 
Application of this PSF produces mean contrasts of 13.0\,\% and 11.3\,\% for the 
small and the large FOV, respectively. 
We now consider the error margins for the non-ideal Voigt contribution to the PSF as 
given in Table~4 in Paper~I. 
First we remain with $\gamma = 0\,\farcs004$ and shift $\sigma$ according to the 
uncertainties of the linear regression stated in Table~4 in Paper~I.
The resulting mean contrast range from 12.9\,\% to 13.4\,\% for the small FOV to 
11.2\,\% and 11.8\,\% for the large FOV. 
The spread in possible contrast values becomes larger if also the uncertainty in the 
parameter $\gamma$ is considered. 
Within the specified error ranges, we can produce contrasts up to 13.2\,\% or even 13.6\,\%.
The intrinsic uncertainty in the PSF apparently allows for the contrasts of the 
degraded synthetic maps in the range of $\sim 11.2$\,\% to $~\sim 13.6$\,\%. 
The observational result of 12.8\,\% is well included in this range. 
In return, the uncertainty in the non-ideal PSF contribution would also lead to 
corresponding error margins for the deconvolved images so that also the original 
synthetic images and the deconvolved observations agree in terms of contrast within 
their mutual error margins.

\paragraph{Limb observations.}  
\label{sec:disclimbobs}
The solar limb is used to re-determine the coordinates of the FOV whenever it can be 
seen in an image. 
The derived positions sometimes differ from the Hinode coordinates in the 
image file headers by of the order of 1\arcsec\ or less. 
At the limb, a small pointing error can translate into a significant 
uncertainty in $\mu$ (e.g.: $\Delta \mu \approx 0.05$ for 1\arcsec).
The unavoidable smearing due to the limited spatial resolution of an  
instrument limits the accuracy of the limb determination. 
The uncertainty can be given as $\Delta \mu < 0.03$ for the 
BFI continuum bands, when assuming one FWHM of the PSF core as maximum offset.
Effects like thermal deformation of the instrument structure and fluctuations in the 
flight characteristics of the spacecraft induce an additional jittering. 
\citet{2007SoPh..243....3K} state that the image stabilization system of SOT achieves 
0\,\farcs007 ($3\,\sigma$ level) and effectively restricts the rms image displacement to 
less than 0\,\farcs03 \citep{2007arXiv0711.1715T, 2007SoPh..243....3K}. 
The influence of the spacecraft on the spatial resolution should thus be rather 
insignificant. 
We nevertheless explicitly state that data for $\mu < 0.1$, corresponding to 
distances less than $\sim 5$\,\arcsec\ from the limb, 
has to be regarded with great care in view of the potential error margins.

\paragraph{Power excess and JPEG compression.}
\label{sec:jpeg}
In Sect.~\ref{sec:decon}, we mentioned a small power contribution at spatial 
scales between the granulation peak and the spatial resolution of SOT 
(see Fig.~\ref{fig:power}).  
\citet{2008A&A...488..375W} describe a similar power excess between SOT and DST 
data (see their Fig.~5). 
It is clearly in excess compared to synthetic images and also is 
usually not seen in comparable SST data. 
We therefore argue that the power excess is an artefact. 
As the enhanced power peak occurs at frequencies that correspond to roughly 
8~pixels and less (see Fig.~\ref{fig:power}), it could in principle be related to 
JPEG compression.  
Essentially all SOT images are compressed before downlinking to the ground  
in order to increase the telemetry throughput. 
The used lossy JPEG image compression uses a $8\,\times\,8$ pixel blocking. 
Consequently artefacts can be expected on pixel scales of 8 and harmonics thereof. 
We therefore compressed and decompressed synthetic images with routines kindly 
provided by Katsukawa \& Tsuneta (priv. comm.). 
The power spectral density of the resulting processed images do not significantly 
differ from those of the original images, disqualifying the JPEG compression as 
the source. 
Whatever the cause of the artefact, 
the effect on the intensity contrast is negligible as the affected scales do not 
contribute much to the overall contrast.  

\section{Conclusion}
\label{sec:conc}

State-of-the-art radiation (magneto-)hydrodynamic simulations can  
reproduce many properties of the observed continuum intensity distribution 
if the effects of instrumental image degradation are properly taken into 
account. 
The latter can have dramatic effects. 
Although SOT is an excellent instrument, which works close to the 
diffraction limit, the non-ideal component to the PSF has a significant effect on 
the intensity distribution.  
The removal of the degradation by PSF-deconvolution more than doubles the 
intensity contrast. 
In contrast to earlier studies, we derive final contrast values at blue, green, 
and red continuum wavelengths that are slightly higher  than for the original 
synthetic images, although they still agree within the mutual variation ranges.  
A detailed realistic treatment of instrumental straylight is found to be of 
crucial importance for a quantitative comparison between simulations and 
observations. 
A constant straylight contribution alone is not sufficient and misleading. 

Owing to the many sources of uncertainty, the intensity contrast as a single 
number is a poor mean of comparing observations and simulations. 
This is already problematic for comparisons of observations that are taken with 
different instruments or under different conditions. 
The only practicable way is to properly remove the effects of image degradation
prior to comparison as it is demonstrated here for SOT. 
This procedure is unfortunatly not always possible as the properties of the PSF 
are often poorly known. 
However, even in such rather fortunate cases as for SOT, one cannot expect an 
exact match between observations and simulations.  
Already the uncertainties in the non-ideal PSF contributions, result in a error 
margins of the order of 1 or possibly even 2 percent points in the blue continuum. 

Comparisons should therefore be based on more informative properties of the 
intensity distribution.  
Here we find the intensity histograms and the power spectral density from 
observation and simulation to agree very well. 
It is obvious that all spatial scales with significant contributions to the 
contrast are resolved with the 50-cm aperture of SOT. 
Therefore we do not expect that observations with larger aperture telescopes 
would yield different contrast values for the quiet Sun. 
We finally conclude that the traditionally perceived conflict between observations 
and simulations can now be dismissed. 

\begin{acknowledgements}
  The authors thank M.~Carlsson, M.~Steffen, F.~W{\"o}ger, H.-G.~Ludwig, O.~Steiner, 
  {\O}.~{Langangen}, S.~Tsuneta, and O.~von der L{\"u}he
  for helpful discussions. 
  S.~Haugan and T.~Fredvik are acknowledged for their support with the Hinode 
  data centre. 
  This work was supported by the Research Council of Norway, grant 
  170935/V30, and a Marie Curie Intra-European Fellowship of the European Commission 
  (6th Framework Programme, FP6-2005-Mobility-5, Proposal No. 042049).   
  Intensive use was made of the Hinode Science Data Centre Europe hosted 
  by the Institute of Theoretical Astrophysics of the University of Oslo, Norway.
  Hinode is a Japanese mission developed and launched by ISAS/JAXA, collaborating 
  with NAOJ as a domestic partner, NASA and STFC (UK) as international partners. 
  Scientific operation of the Hinode mission is conducted by the Hinode science team 
  organized at ISAS/JAXA. This team mainly consists of scientists from institutes in 
  the partner countries. Support for the post-launch operation is provided by JAXA 
  and NAOJ (Japan), STFC (U.K.), NASA, ESA, and NSC (Norway). 
  This research has made use of NASA's Astrophysics Data System.
\end{acknowledgements}
\bibliographystyle{aa}

\end{document}